\documentclass[float,epsfig,usenatbib]{mnras}

\usepackage{graphicx}
\usepackage{appendix}
\usepackage{url}
\usepackage{psfrag}
\usepackage{amsmath,amssymb}
\usepackage{threeparttable}
\usepackage{float}
\usepackage{subfigure}
\usepackage{afterpage}
\def\apj{ApJ}
\def\apjl{ApJL}
\def\mnras{MNRAS}

\def\pasp{PASP}
\def\aj{AJ}

\def\aap{A\&A}
\def\apjs{ApJS}

\title[Quenching vs. Structural Transformation]{The SAMI Galaxy Survey: Satellite galaxies 
undergo little structural change during their quenching phase}
\author[L. Cortese et al.]
{L. Cortese\thanks{luca.cortese@uwa.edu.au}$^{1,2}$, J. van de Sande$^{2,3}$, C.~P. Lagos$^{1,2}$, B. Catinella$^{1,2}$, L.~J.~M. Davies$^{1}$,\newauthor S.~.M. Croom$^{2,3}$, S. Brough$^{2,4}$, J.~J. Bryant$^{2,3,5}$, J.~S. Lawrence$^{6}$, M.~S.~Owers$^{7}$,\newauthor S.~N. Richards$^{8}$, S.~M.~Sweet$^{2,9}$\\ 
$^1$International Centre for Radio Astronomy Research, The University of Western Australia, 35 Stirling Hw, 6009 Crawley, WA, Australia\\
$^2$ARC Centre of Excellence for All Sky Astrophysics in 3 Dimensions (ASTRO 3D)\\
$^3$Sydney Institute for Astronomy, School of Physics, The University of Sydney, Sydney, New South Wales, Australia\\
$^4$School of Physics, University of New South Wales, NSW 2052, Australia\\
$^5$Australian Astronomical Optics, AAO-USydney, School of Physics, University of Sydney, NSW 2006, Australia\\
$^6$Australian Astronomical Optics - Macquarie, Macquarie University, NSW 2109, Australia\\
$^7$Department of Physics and Astronomy, Macquarie University, NSW 2109, Australia\\
$^8$SOFIA Science Center, USRA, NASA Ames Research Center, Building N232, M/S 232-12, P.O. Box 1, Moffett Field, CA 94035-0001, USA\\
$^9$Centre for Astrophysics and Supercomputing, Swinburne University of Technology, PO Box 218, Hawthorn, VIC 3122, Australia\\
}
\date{}
\begin{document}
\newcommand{\Zsolar}{\mbox{$\,\rm Z_{\odot}$}}
\newcommand{\Msolar}{\mbox{$\,\rm M_{\odot}$}}
\newcommand{\Lsolar}{\mbox{$\,\rm L_{\odot}$}}
\newcommand{\xs}{$\chi^{2}$}
\newcommand{\dxs}{$\Delta\chi^{2}$}
\newcommand{\xsn}{$\chi^{2}_{\nu}$}
\newcommand{\ls}{{\tiny \( \stackrel{<}{\sim}\)}}
\newcommand{\gs}{{\tiny \( \stackrel{>}{\sim}\)}}
\newcommand{\asec}{$^{\prime\prime}$}
\newcommand{\amin}{$^{\prime}$}
\newcommand{\mstar}{\mbox{$M_{*}$}}
\newcommand{\hi}{H{\sc i}}
\newcommand{\hii}{H{\sc ii}\ }
\newcommand{\kms}{km~s$^{-1}$\ }

\maketitle

\label{firstpage}

\begin{abstract}
At fixed stellar mass, satellite galaxies show higher passive fractions than centrals, suggesting that environment is directly quenching their star formation. 
Here, we investigate whether satellite quenching is accompanied by changes in stellar spin (quantified by the ratio of the rotational 
to dispersion velocity V/$\sigma$) for a sample of massive ($M_{*}>$10$^{10}$ M$_{\odot}$) satellite galaxies extracted from the SAMI Galaxy Survey. 
These systems are carefully matched to a control sample of main sequence, high $V/\sigma$ central galaxies.    
As expected, at fixed stellar mass and ellipticity, satellites have lower star formation rate (SFR) and spin than the control centrals. However, most of the difference is in SFR, whereas the spin decreases significantly only for satellites that have already reached the red sequence. We perform a similar analysis for galaxies in the EAGLE hydro-dynamical simulation and recover differences in both SFR and spin similar to those observed in SAMI. However, when EAGLE satellites are matched to their {\it true} central progenitors, the change in spin is further reduced and galaxies mainly show a decrease in SFR during their satellite phase. The difference in spin observed between satellites and centrals at $z\sim$0 is primarily due to the fact that satellites do not grow their angular momentum as fast as centrals after accreting into bigger halos, not to a reduction of $V/\sigma$ due to environmental effects. Our findings highlight the effect of progenitor bias in our understanding of galaxy transformation and they suggest that satellites undergo little structural change before and during their quenching phase.                 
\end{abstract}

\begin{keywords}
galaxies: evolution--galaxies: fundamental parameters--galaxies: kinematics and dynamics
\end{keywords}

\section{Introduction}
Observational evidence that galaxy properties vary as a function of environment has been 
presented since at least \cite{HUBH31}. After almost a century, it is now clear that the structure 
(usually quantified via visual classification or two-dimensional surface brightness decomposition) and 
star formation activity of galaxies depend on their location within the large scale structure (e.g., \citealp{dressler80,lewis02,gomez03,review,wetzel12}). 

It is also firmly established that these trends, generally referred to as `morphology-density' and `star formation rate-density' 
relations, are not simply two different manifestations of the same evolutionary paths. For example, there is plenty of evidence for the existence of a large population of rotationally-supported, disky systems with low (or no) star formation in groups and clusters (e.g., \citealp{vandberg76,poggianti99,ha06,lisker06,dEale,bamford09,cortese09,toloba09,bundy10,hester10}). 
Thus, separating between quenching and structural transformation becomes critical to reveal what shaped the environmental trends that we see today. 

The advent of large-area spectroscopic surveys and the refinement of large-scale cosmological simulations have also highlighted that the way 
in which we define `environment' does matter (e.g., \citealp{muldrew12,fossati15}). There is no `golden environmental ruler', every metric has its advantages and disadvantages and the definition of environment should be tuned to the particular issue being addressed (e.g., \citealp{brown17}). Nevertheless, it is now well established that one of the best ways to isolate galaxies most likely to be affected by environment is to focus on satellites. Central galaxies dominate in number at all stellar masses (e.g., \citealp{tempel09,yang09}), and it is still debated whether or not their evolution is significantly affected by environment (e.g., \citealp{blanton07,vdbosch08,willman10}). Thus, including centrals in the analysis would significantly reduce or completely wash out any signatures of environmentally-driven transformation. 

Interestingly, while the focus on satellite galaxies has reduced the disagreement between some observational results, this approach turns out 
not to be sufficient to separate the relative importance of quenching and morphological transformation in the life of satellite galaxies. Indeed, observational evidence supporting seemingly opposite transformation scenarios has been presented, namely simultaneous quenching and morphological transformation on one side (e.g., \citealp{moss00,christlein04,cappellari13,george13,omand14,Kawinwanichakij17}), 
and quenching-only followed by no or minor structural transformation on the other (e.g., \citealp{LARS80,blanton05,cortese09,woo17,rizzo18}). 
There are various potential reasons behind these 
conflicting results, but our view is that most of the difference can be ascribed to two - equally important - limitations. 

First, the techniques used to quantify structure/morphology vary significantly in the literature, encompassing both visual classification (generally used to isolate early- from late-type galaxies) and structural parameters obtained via two-dimensional surface brightness decomposition of optical images. Arguably, neither of the two has a direct connection to the kinematic properties of galaxies, as it has now been demonstrated that they are not able to distinguish between rotationally- and dispersion-supported systems (e.g., \citealp{emsellem11,krajnovic13,cortese16}). 
Thus, to identify and quantify truly structural transformation, and separate it from visual changes simply due to quenching and disk fading, information 
on the kinematic properties of stars is vital. 

Second, it is now well established that, for massive satellite galaxies (stellar masses $M_{*}>$10$^{10}$ M$_{\odot}$), full quenching takes at least a few Gyrs after infall (e.g., \citealp{cortese09,weinmann10,wetzel13,oman16,bremer18}), a time during which central star-forming systems have grown significantly \citep{vanderwell14}. This means that today's centrals cannot be naively assumed to be representative of the progenitor population of local satellites and used to quantify the effect of nurture on galaxy evolution, an issue generally refereed to as {\it progenitor bias} \citep{vandokkum01,woo17}. Only by identifying the real progenitors of satellites at the time of infall we can reveal how satellites have been transformed by environment. While this is still out of reach from an observational perspective, the improvement of cosmological simulations is starting to make it possible to use models to quantify the effect of {\it progenitor bias} and try to correct for it. 

In this paper, we revisit the issue of satellite transformation with the goal of quantifying the change in star formation activity and structure {\it separately}, and to determine 
if they both happen simultaneously or on different time-scales. Our analysis improves on previous works by directly addressing the two limitations discussed above. 
First, we take advantage of optical integral field spectroscopic observations obtained as part of the SAMI Galaxy Survey \citep{bryant15} to directly trace the stellar kinematic of galaxies. Second, we compare our findings with predictions from the Evolution and Assembly of GaLaxies and their Environments (EAGLE; \citealp{eagle15}) cosmological simulation, and use it to quantify the effect of {\it progenitor bias}.
The use of a cosmological simulation such as EAGLE turns out to be critical for a less biased interpretation of SAMI data, highlighting the danger of inferring galaxy evolutionary histories from single-epoch snapshots.

This paper is organized as follows. In Sec. 2 we describe how our sample is extracted from the SAMI Galaxy Survey, 
the stellar kinematic parameters as well as the ancillary data used in this paper. In Sec. 3, we compare the star formation 
and kinematic properties of satellites and centrals and compare our results with the prediction from the EAGLE simulation. 
This section includes the main results of this work. Lastly, the implications of our results are discussed in Sec. 4. 

Throughout this paper, we use a flat $\Lambda$ cold dark matter concordance 
cosmology: $H_{0}$ = 70 km s$^{−1}$ Mpc$^{−1}$, $\Omega_{0}$=0.3, $\Omega_{\Lambda}$=0.7.

\section{The data}
\label{QC}
The SAMI Galaxy Survey has observed $\sim$3000 individual galaxies in the redshift 
range 0.004$<z<$0.095 and with stellar masses greater than $\sim$10$^{7.5}$ M$_{\odot}$ taking advantage of the Sydney-AAO Multi-object Integral field spectrograph (SAMI; \citealp{croom12}), installed at the Anglo-Australian Telescope.
SAMI is equipped with photonic imaging bundles (`hexabundles', \citealp{bland11,hexa14}) 
to simultaneously observe 12 galaxies across a 1 degree field of view.
Each hexabundle is composed of 61 optical fibres, each with a diameter of $\sim$1.6\arcsec, 
covering a total circular field of view of $\sim$14.7\arcsec\ in diameter. 
SAMI fibres are fed into the AAOmega dual-beam spectrograph, providing 
a coverage of the 3650-5800 \AA\ and 6240-7450 \AA\ wavelength ranges with dispersions 
of 1.05 \AA\ pixel$^{-1}$ and 0.59 \AA\ pixel$^{-1}$, respectively. 
\begin{figure*}
\centering
\includegraphics[width=16.cm]{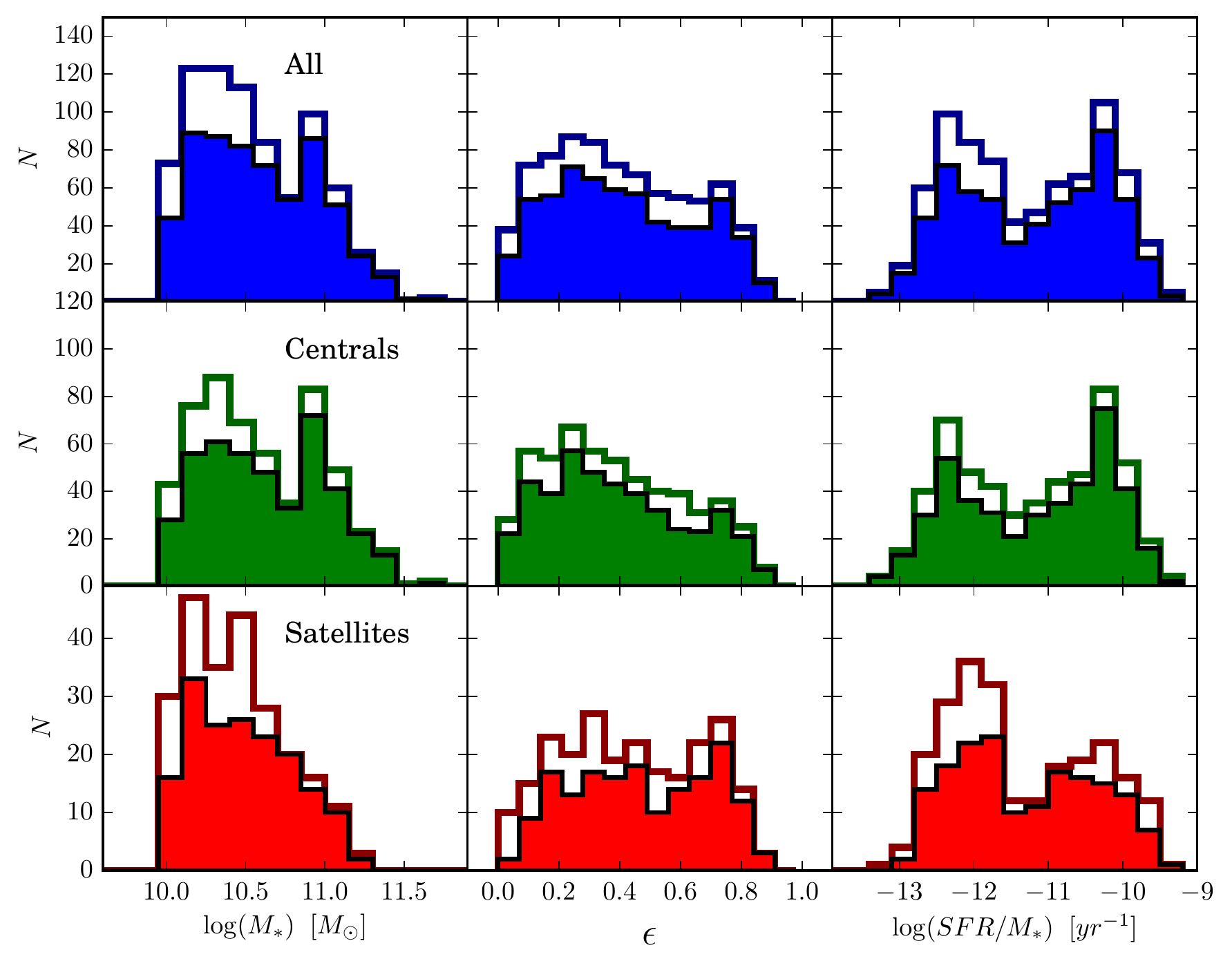}
\caption{The stellar mass ($M_{*}$, left), ellipticity ($\epsilon$, middle) and specific star formation rate ($SFR/M_{*}$, right) distribution 
for our {\it parent} (empty histogram) and {\it final} (filled histogram). The top row includes all galaxies, while the middle and bottom rows focus on central and satellite galaxies only. It is clear that our final sample covers the same parameter space as our initial parent sample.}
\label{diagnostic}
\end{figure*}

In this paper, we extract our sample from the 1552 galaxies overlapping with the footprint of the Galaxy And Mass Assembly Survey (GAMA, \citealp{gama}) 
included in the SAMI public data release 2 \citep{scott2018} and for which integrated current star formation rate (SFR) estimates are 
available (referred to as {\it parent sample}). SFRs are taken from \cite{davies16} and have been 
derived by fitting the spectral energy distribution fitting code MAGPHYS \citep{dacunha08} to the full 
21-band photometric data available for GAMA galaxies across the ultraviolet to the far-infrared 
frequency range \citep{driver16,wright16}. In addition to the wealth of multi-wavelength data available, the GAMA 
regions are characterised by an exquisitely high spectroscopic completeness, providing 
us with a state-of-the-art group catalogue \citep{robotham11}, critical for distinguishing between central and satellite galaxies. 

We focus on galaxies with stellar mass greater than 10$^{10}$ M$_{\odot}$ (768 galaxies), 
for which the signal-to-noise (S/N) in the continuum is generally high enough to allow a proper 
reconstruction of the stellar velocity field. Stellar masses ($M_{*}$) are estimated from $g-i$ colours and $i$-band 
magnitudes following \cite{taylor11}, as described in \cite{bryant15}.

The procedure adopted to extract stellar kinematic parameters is extensively described 
in \cite{vds17} and \cite{scott2018}. Here, we briefly summarise its key steps.  
Stellar line-of-sight velocity and intrinsic dispersion maps are obtained using 
the penalised pixel-fitting routine \textsc{ppxf}, developed by \cite{cappellari2004}.
SAMI blue and red spectra are combined by convolving the red spectra to match the instrumental 
resolution in the blue. We then use the 985 stellar template spectra from the MILES stellar library \citep{miles} 
to determine the best combination of model templates able to reproduce the galaxy 
spectrum extracted from annular binned spectra following the optical ellipticity and position angle of the 
target. We apply the following quality cuts to each spaxel to discriminate between good and bad fits \citep{vds17}: 
$S/N>$3 \AA$^{-1}$, $\sigma>FWHM_{instr}$/2$\sim$35 km s$^{-1}$, $V_{err}<$30 km s$^{-1}$ and $\sigma_{err}<$ $\sigma\times$0.1 + 25 km s$^{-1}$, 
where $V$, $V_{err}$, $\sigma$ and $\sigma_{err}$ are the line-of-sight and dispersion velocities and their uncertainties. 

The ratio of ordered versus random motions $V/\sigma$ within one effective radius is then determined as in \cite{cappellari07}: 
\begin{equation}
\Big(\frac{V}{\sigma}\Big)^{2} = \frac{\sum F_{i}V_{i}^{2}}{\sum F_{i}\sigma_{i}^{2}} 
\end{equation}
where $F_{i}$ is the flux in each spaxel. We sum only spaxels included within an ellipse 
of semi-major axis corresponding to one effective radius in r-band and position angle and ellipticity 
taken from v09 of the GAMA single S\'ersic profile fitting catalogue \citep{kelvin12}. 
We require that at least 95\% of the spaxels within the aperture full fill our quality cuts to flag 
the estimate of $V/\sigma$ as reliable. This reduces our sample to 726 galaxies. 

As SAMI galaxies cover a wide range of effective radii, we want to make sure that the one effective 
radius aperture provides a reasonable number of independent resolution elements to determine $V/\sigma$, and minimise 
the effect of beam smearing.
Thus, we remove all galaxies with $r_{e}<$2\arcsec or $r_{e}$ smaller than 2.5 the half-width at 
half-maximum of the point-spread-function of the secondary standard star observed with the same plate (121 galaxies).
Conversely, we keep galaxies with effective radii larger than the SAMI bundle (154 objects) and apply the 
aperture correction as described in \cite{vds17b} to recover the value of $V/\sigma$ within one effective radius.

\begin{figure*}
\centering
\includegraphics[width=18.cm]{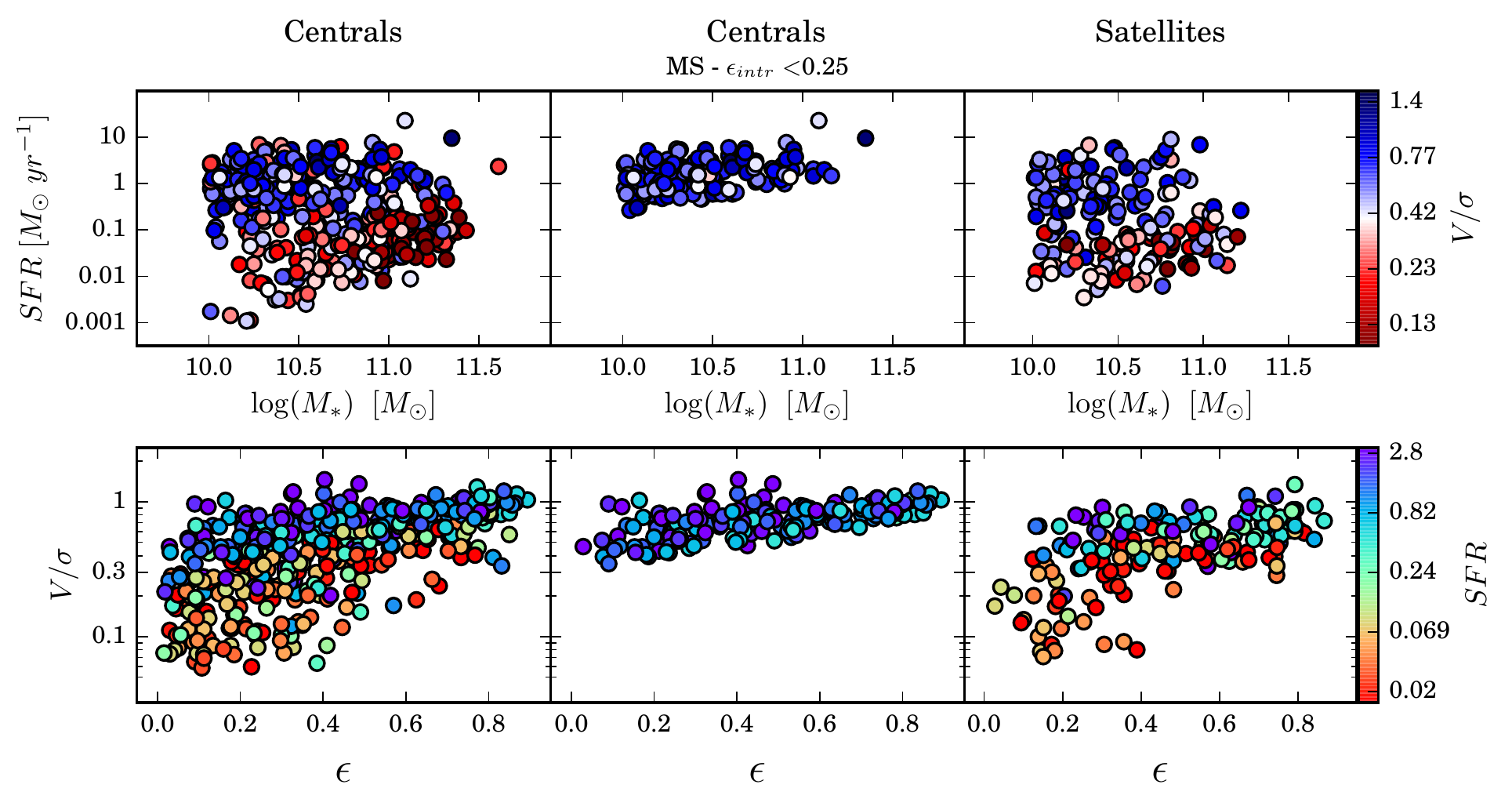}
\caption{The M$_{*}$-SFR (top) and $V/\sigma$-$\epsilon$ (bottom) planes for all centrals in our sample (left), main-sequence, disky centrals (middle) and satellite galaxies (right). Points are colour-coded by $V/\sigma$ and $SFR$ in the top and bottom panels, respectively.}
\label{sample}
\end{figure*}

The selections described above reduce our sample from 768 to 605 galaxies. During the analysis described in Sec. 3, five satellite 
galaxies were further removed from the sample as visual inspection highlighted issues with their photometric ellipicity and/or position angles (e.g., contamination by foreground/background objects, structural parameters tracing the inner bar instead of the disk, etc.). 
In conclusion, the {\it final sample} used in this paper is composed of 600 galaxies, 431 of which are centrals and 169 are group satellites according to v09 of the group catalogue by \cite{robotham11}. Our satellites occupy halo masses up to $\sim10^{14.5}$ M$_{\odot}$, with an average value of $\sim$10$^{13.4}$ M$_{\odot}$. 

The stellar mass, $r$-band ellipticity ($\epsilon$) and specific star formation rate distributions for our {\it parent} and {\it final} samples are shown in Fig.~\ref{diagnostic} as empty and filled histograms, respectively. All galaxies are shown in the top, with only centrals/satellites included in the middle/bottom row, respectively. It is clear that our quality cuts preferentially affect round, low-mass, passive objects. However, as the two samples cover the same parameter space in all three variables, we are confident that the matching procedure at the basis of our analysis in Sec. 3 is not biased by the strict criteria used to extract our {\it final} sample. Indeed, our main conclusions and average trends are not affected even if we relax the criteria used to exclude `marginally resolved' galaxies, with the only noticeable change being an increase in scatter. 
The potential effect of beam smearing on our estimates of $V/\sigma$ is discussed in Appendix \ref{appendix}, where we show that correcting for beam smearing 
would even reinforce the main conclusions of this paper.

\section{Quenching and structural transformation at $z\sim$0}
Our primary goal is to {\it separately} quantify the changes in $SFR$ and $V/\sigma$ (a proxy for the stellar spin parameter) experienced by 
galaxies after they have become satellites. As shown in Fig.~\ref{sample}, and consistently with previous works (e.g., \citealp{vdbosch08,weinmann09,peng12}), the fraction of satellite galaxies with low specific star formation rate in our parent sample is significantly larger than that of centrals. This supports the common assumption that environmental effects are playing a more active role in the 
evolution of satellite than in centrals. Our aim is to determine if satellites being quenched after infall do also experience changes in their kinematic properties. 

Ideally, this would require a-priori knowledge of the properties of satellite galaxies at the time of infall 
into their host halo. While this is currently possible in cosmological simulations, observationally we are not yet able to link progenies and 
progenitors at different redshifts. Thus, nearly all observational studies so far have used central galaxies at $z\sim$0 to `guess' the properties of galaxies 
at the time when they became satellites (e.g., \citealp{vdbosch08,woo17}). 

In this work, we first make a similar assumption to quantify the variation in stellar kinematic between SAMI satellites and centrals. 
We then compare our results with the predictions of the EAGLE hydrodynamical simulation \citep{eagle15,crain15,mcalpine16} at $z\sim$0. 
This is needed to validate the ability of the simulation to reproduce the observed difference between centrals and satellites. 
Lastly, we use EAGLE to quantify the effect of progenitor bias on the $z\sim$0 comparison. 
This last step is the most critical one for the interpretation of the results emerging from the SAMI data.

\subsection{SAMI galaxies}
\label{SAMImatch}
In order to quantify the amount of transformation experienced by SAMI satellites, we compare their properties to those of rotationally-supported centrals in the star-forming main sequence. 
Of course, this is very conservative and would imply that all galaxies become satellites as rotating star-forming disks. As we have evidence 
that this is not always the case (e.g., \citealp{big,mei07}), our findings must be interpreted as an upper limit for the real amount of transformation 
experienced by galaxies during their satellite phase. We will further discuss this point in the following sections.

We isolate star-forming centrals by selecting systems with SFR higher than the lower 1$\sigma$ envelope of the $z\sim$0 main sequence obtained by \cite{davies16} for 
GAMA galaxies, namely $\log(SFR)>0.7207\times(\log(M_{*}/M_{\odot})-10) +0.061 -0.73$. Similarly, rotationally-supported centrals are selected by imposing that 
$\log(V/\sigma)> 0.4\times\epsilon -0.5$, where $\epsilon$ is the observed ellipticity in $r$-band. Following the formalism in \cite{cappellari16}, this is nearly equivalent to selecting only axisymmetric galaxies with intrinsic ellipticity ($\epsilon_{intr}$) smaller than $\sim$0.25 for anisotropy $\beta_{z}$=0.6$\times\epsilon_{intr}$, i.e., consistent with what observed for disk-dominated galaxies (e.g., \citealp{giovanelli94,Unterborn08,foster17}). We favour this empirical criterion to the analytical prescription as it provides a more conservative cut at low ellipticities, where the difference between the analitic prescription for different intrisic shapes becomes significantly smaller than the measurement errors in both $V/\sigma$ and $\epsilon$. 
The combination of both criteria yields 167 star-forming, rotating centrals, including both isolated (i.e., with no detected companions: 90 objects) and group centrals, with the vast majority of centrals in groups (55 out of 77 objects) having just one or two satellites according to the GAMA group catalog.

The results of our selection are shown in Fig.\ref{sample}, where we compare the distribution in the $SFR$-$M_{*}$ (top row) and $V/\sigma-\epsilon$ plane (bottom row) of all 431 centrals in our sample (left column), and for the 167 main-sequence, rotationally-supported centrals (middle column). For reference, we also show the distribution of the 169 group satellites (right column).
Points are colour-coded by $V/\sigma$ in the $SFR$-$M_{*}$ plane (top row) and $SFR$ in the $V/\sigma-\epsilon$  (bottom row) to highlight the tight apparent link between SFR and $V/\sigma$.
Fig.~\ref{sample} also shows that the $V/\sigma$ and $SFR$ cuts adopted to isolate our control sample of central galaxies are equivalent from a statistical point of view: i.e., applying only one of the two would result in a control sharing the same properties and, indeed, would lead us to the same results. This is also consistent with the tight correlation between $V/\sigma$ and stellar age recently presented by \cite{jesse18}. The simultaneous use of the two cuts is preferred simply because it provides a more rigorous initial hypothesis to our exercise (i.e., it gives independent constraints to both star formation activity and structural properties of the control sample).

In order to quantify the difference in SFR and spin of satellites compared to main-sequence, rotationally-supported centrals, we follow a technique similar to that discussed in \cite{ellison15,ellison18}. We define $\Delta (SFR)$ and $\Delta (V/\sigma)$ as the difference (in log-space) between the SFR or $V/\sigma$ of a satellite and the median value obtained for a control sample of main-sequence, rotation-dominated centrals matched in both stellar mass and ellipticity. During the matching procedure, we 
start isolating all the control centrals within 0.15 dex in stellar mass and 0.1 in ellipticity from each satellite. If such control sample includes fewer than 10 galaxies, we iteratively increase the range of stellar mass and ellipticity (in steps of 0.01) until the control includes at least 10 objects. The end result is that our average bins are $\sim$0.16 dex and 0.11 wide in stellar mass and ellipticity, respectively. We then compute the median SFR and $\epsilon$ for the control and use it to determine $\Delta (SFR)$ and $\Delta (V/\sigma)$ for each satellite. 

The additional matching by ellipticity is adopted mainly because the $V/\sigma$ estimates do not include an inclination correction. This is also justified by the fact that the ellipticity distribution of central and satellites may not always be the same (e.g., see Fig.~\ref{sample}). The fact that, for SAMI galaxies, observed and intrinsic ellipticity do not correlate \citep{jesse18} also suggests that this assumption is not introducing any significant bias. Indeed, matching only by stellar mass would not change our results. Our findings are also unchanged if we limit our control sample to isolated or group centrals only.

It is important to acknowledge that, despite some differences in the technique used here, our quantification of $\Delta (SFR)$ and $\Delta (V/\sigma)$ is deeply inspired by the definition of atomic gas (\hi) deficiency originally introduced by \cite{haynes}. By quantifying the difference in \hi\ content with respect to galaxies of same morphology and size, \hi\ deficiency has become a key parameter for isolating the effect of environment on the cold gas content of galaxies (e.g., \citealp{giova85,solanes01,review,cortese11,cortese16b}).

\begin{figure}
\centering
\includegraphics[width=8.5cm]{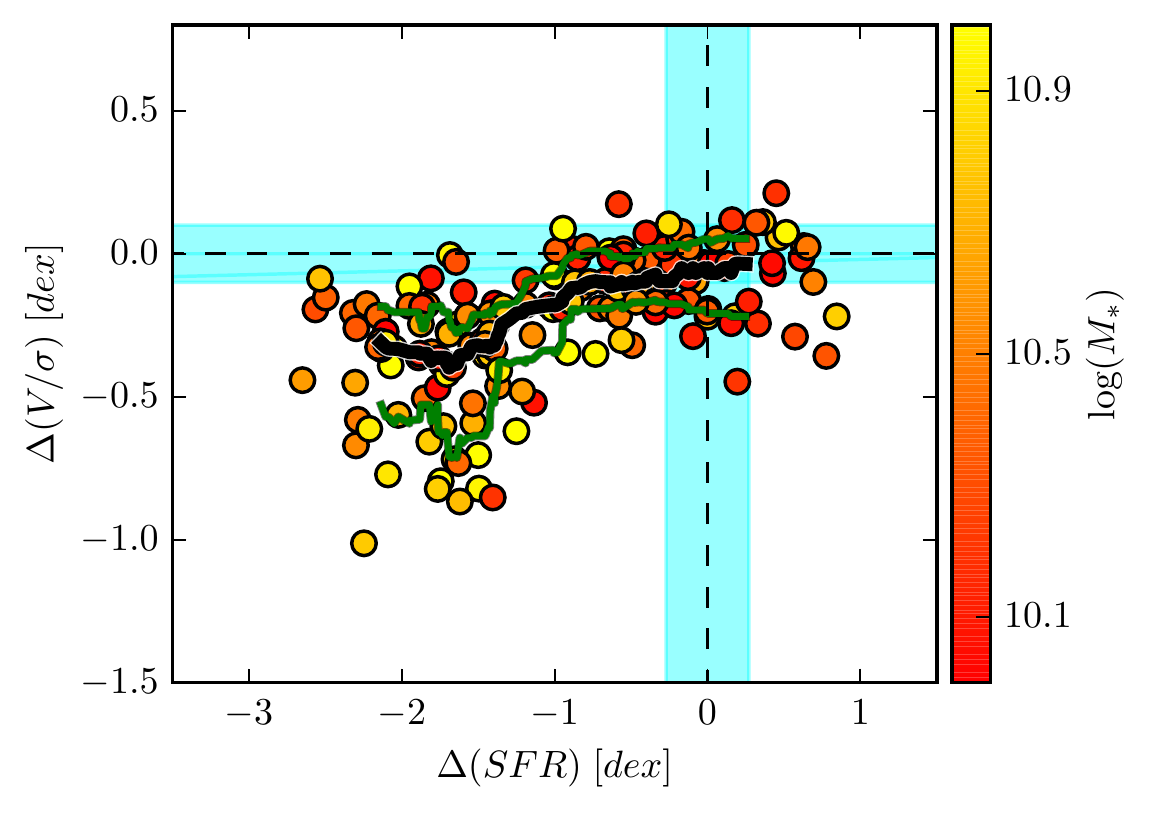}
\caption{Variations in stellar $V/\sigma$ and $SFR$ for satellite galaxies with respect to our control sample of main-sequence, high $V/\sigma$ centrals. Points are colour-coded by stellar mass. 
Dashed lines and cyan bands show the average and standard deviation for the control sample. The thick black and thin green lines show the running median and 20\%-80\% percentile ranges for $\Delta (V/\sigma)$ in bins of $\Delta (SFR)$. See \S~\ref{SAMImatch} for details on the matching procedure.}
\label{delta}
\end{figure}

In Fig.~\ref{delta}, we show the result of the matching procedure by plotting $\Delta (V/\sigma)$ vs. $\Delta (SFR)$, with points colour-coded by stellar mass. Dashed lines define `normalcy' (i.e., no change) in SFR and/or $V/\sigma$, with cyan bands highlighting the 1$\sigma$ variation for the control sample. If satellites were to first loose spin and then decrease their star formation with respect to centrals, they would move vertically downwards (i.e., negative $\Delta (V/\sigma)$ around $\Delta (SFR)\sim$0) and then horizontally towards the left (negative $\Delta (SFR)$ and negative $\Delta (V/\sigma)$). Similarly, if changes in $SFR$ were followed by similar changes in stellar spin, satellites would form a diagonal sequence showing $\Delta (SFR)\propto\Delta (V/\sigma)$. Conversely, satellite galaxies occupy an L-shaped parameter space in the $\Delta (SFR)$-$\Delta (V/\sigma)$ plane with large changes in $V/\sigma$ only for the passive population.

Main-sequence satellite galaxies show an average $V/\sigma$ marginally lower than that of our control sample ($\Delta (V/\sigma)$ $\sim$$-$0.08, with standard deviation $\sim$0.13 dex). 
During the satellite quenching phase, $\Delta (V/\sigma)$ remains roughly constant until galaxies have reduced their current star formation rate by more than a factor of ten. Then, 
for the more passive population ($\Delta (SFR)\sim -$1.8 dex), the scatter in $\Delta (V/\sigma)$ more than doubles and satellites span almost a dex in $\Delta (V/\sigma)$, although the median value never goes below $-$0.4 dex. This is qualitatively consistent with previous observational (e.g., \citealp{review,cortese09,woo17}) and theoretical works (e.g., \citealp{correa17}) suggesting the presence of a wide range of visual and/or photometric morphologies in the red sequence of satellite galaxies.

No significant dependence of the position of satellites in the $\Delta (SFR)$-$\Delta (V/\sigma)$ plane on stellar mass (or group halo mass, not shown here) is observed. Intriguingly, the three outliers in the bottom-right quadrant (i.e., positive $\Delta (SFR)$ and negative $\Delta (V/\sigma$)) are all interacting systems (GAMA ID 301382, 485833, 618992), suggesting that our technique may also be able to identify boosts in $SFR$ accompanied by kinematic perturbations. 

It is tempting to interpret Fig.~\ref{delta} in terms of galaxy transformation, and consider the variation of $\Delta (SFR)$ and $\Delta (V/\sigma)$ as the evolutionary paths followed by satellites after infall. As such, one would immediately conclude that satellites experience a two-phase transformation, with quenching of the star formation happening first and structural transformation - if any - taking place at later stages or on longer time-scales, and visibly affecting only galaxies already quenched. Unfortunately, Fig.~\ref{delta} would directly show evolutionary tracks only if the vast majority of satellite galaxies at $z\sim$0 had become satellites in the last couple of billion years. As this is clearly not the case (e.g., \citealp{delucia12,han18}), their properties at the time of infall could be significantly different from those of central galaxies in the local Universe. Not only their stellar mass was likely smaller, potentially undermining the basis of our matching procedure but, most importantly, their SFR was higher and their spin parameter lower than those of star-forming centrals at $z\sim$0 with the same mass. Thus, our results most likely provide just an upper limit to the change in $V/\sigma$ parameter and a lower limit to the change in SFR experienced by satellite galaxies. We will demonstrate this point in the next section.

%

\subsection{Simulated galaxies in EAGLE}
In order to quantify the potential effect of {\it progenitor bias} on the results presented in Fig.~\ref{delta}, we perform the same analysis presented in the 
previous section on galaxies extracted from the EAGLE simulation. 
We focus on the EAGLE reference model, denoted as Ref-L100N1504 and rescaled to the cosmology adopted in this paper, which corresponds to a cubic volume of 100 comoving Mpc per side, and use the 
stellar kinematic measurements presented in \cite{lagos18}. Briefly, stellar kinematic maps are produced by projecting the stellar particle kinematic 
properties on a two-dimensional plane with bin size of 1.5 comoving kpc. The line of sight is fixed along the z-axis of the simulated box, providing a random distribution for the orientation of galaxies, and line-of-sight and dispersion velocities are obtained by fitting a Gaussian to line-of-sight velocity distribution for each pixel. 
The $V/\sigma$ ratio is then estimated in the same way as in the observations, by integrating only pixels within one effective radius and using the $r$-band luminosity of each pixel as weight. 
Star formation rate is implemented following the prescription of \cite{schaye08}, and here we use total current star formation rates as described in \cite{furlong15}.
Central galaxies in the simulation are defined as those objects hosted by the main subhalo, while galaxies hosted in other subhaloes within the group are considered satellites.
Across the stellar mass range of interest of this paper (10$<\log(M_{*}/M_{\odot})<$11.5), we find 2265 centrals and 1413 satellites. 
Satellite galaxies in EAGLE span a slightly wider range of halo masses than our SAMI {\it final sample}, extending up to $\sim$10$^{14.8}$ M$_{\odot}$ with an average halo mass of $\sim$10$^{13.6}$ M$_{\odot}$: i.e., $\sim$0.2 dex higher than our {\it final sample}.

Because the main sequence of star-forming galaxies in EAGLE is slightly offset towards lower SFR with respect to the observed one \citep{furlong15}, we revise the cut used to isolate star-forming centrals for the matching procedure: i.e., $\log(SFR)>0.7207\times(\log(M_{*}/M_{\odot})-10) +0.061 -1.2$. Similarly, EAGLE passive galaxies have naturally SFR equal to 0, whereas SAMI red-sequence objects have their star formation clustered around $\sim$10$^{-1.8}$ $M_{\odot}$ yr$^{-1}$. This is due to the inability of SED-fitting techniques to quantify very low levels of SFRs. For consistency with observations, EAGLE galaxies with SFR$<$10$^{-1.8}$ $M_{\odot}$ yr$^{-1}$ are assigned a random value of SFR following a log-normal distribution peaked at 10$^{-1.8}$ $M_{\odot}$ yr$^{-1}$, with 0.2 dex scatter. We note that the exact location and shape of the distribution used to re-scale passive galaxies does not affect our results. Our final sample used for the matching is thus composed of 1204 main-sequence, rotationally-supported centrals and 1413 satellites. 
\begin{figure}
\centering
\includegraphics[width=8.5cm]{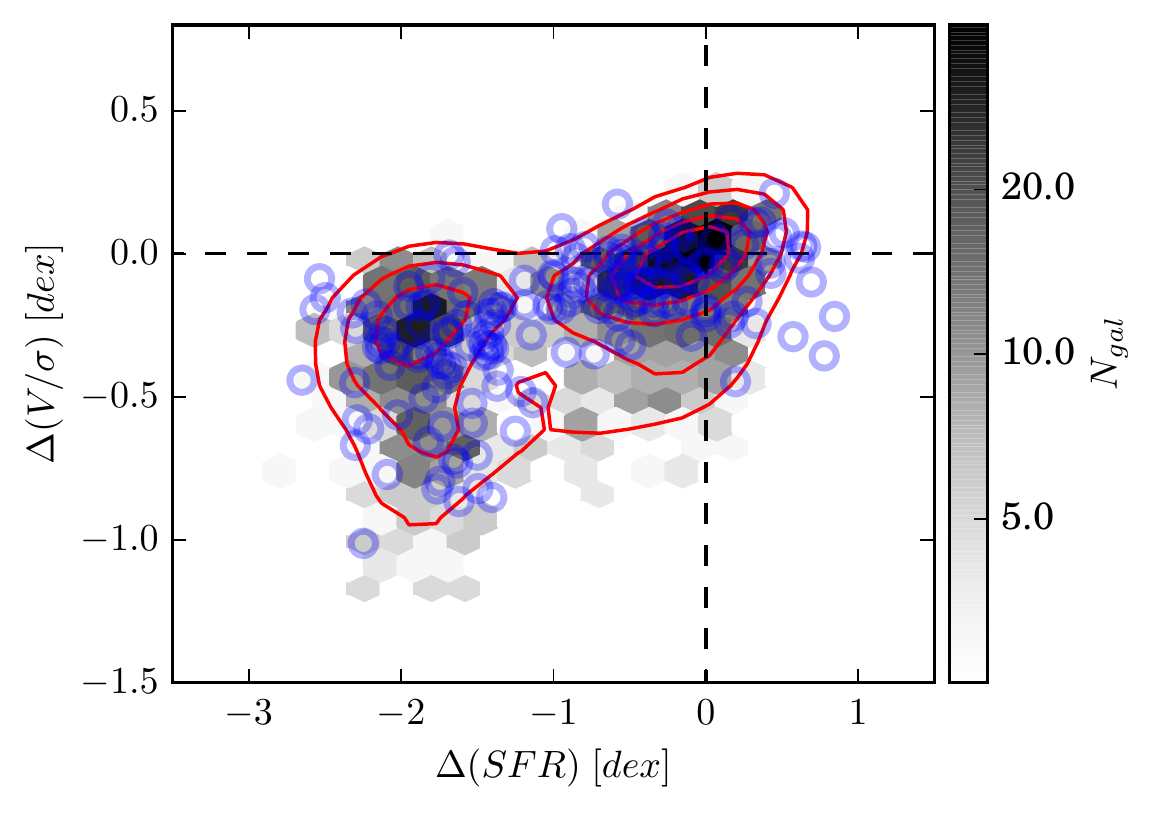}
\caption{Variations in stellar $V/\sigma$ and $SFR$ for satellite galaxies in the EAGLE simulation. Density distribution and Gaussian kernel contours are shown in grey and red, respectively. Matching is done following the same technique used for SAMI galaxies, overplotted as empty blue circles for comparison.}
\label{deltaeagle}
\end{figure}

We perform a matching procedure identical to the one used for SAMI data. Namely, each satellite is matched with all main-sequence, rotationally-supported centrals within 0.15 dex in stellar mass and 0.1 in ellipticity. The median SFR and $V/\sigma$ are then used to estimate $\Delta (SFR)$ and $\Delta (V/\sigma)$ for each satellite. 
The result is shown in Fig.~\ref{deltaeagle}. The density distribution of EAGLE galaxies is highlighted in grey, with Gaussian kernel density contours in red. SAMI galaxies are overplotted as empty blue circles for comparison. We find greement between the distribution of SAMI and EAGLE galaxies, with the values of $\Delta (V/\sigma)$ for EAGLE galaxies becoming large only for galaxies already in the passive population.  

\begin{figure}
\centering
\includegraphics[width=8.5cm]{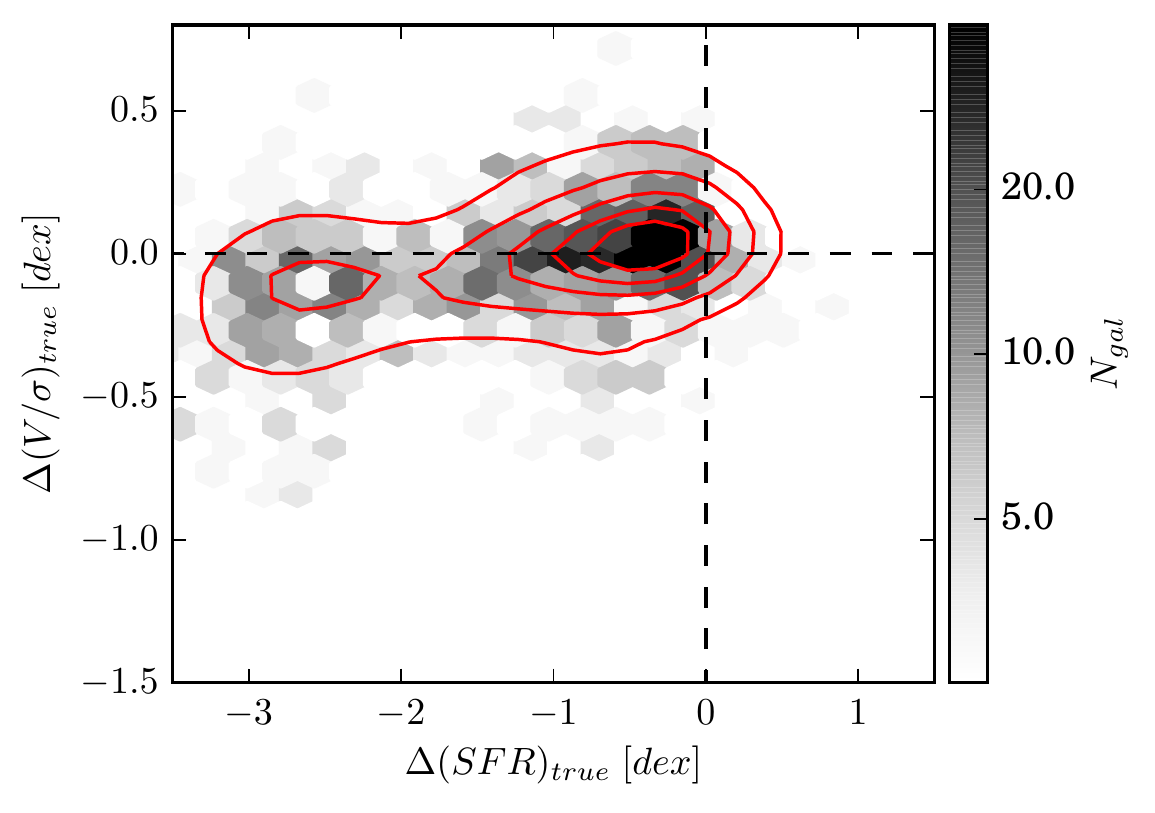}
\caption{`True' variations in stellar $V\sigma$ and $SFR$ for satellite galaxies in the EAGLE simulation, determined by comparing the $z\sim$0 properties to those at the last snapshot during which the galaxy was a central. Density distribution and contours are as in Fig.~\ref{deltaeagle}}
\label{deltareal}
\end{figure}

The good agreement between SAMI and EAGLE gives us confidence to use EAGLE to investigate the effect of {\it progenitor bias} in our analysis. To do so, in Fig.~\ref{deltareal} we plot $\Delta (SFR)_{true}$ vs. $\Delta (V/\sigma)_{true}$, estimated by comparing the satellite's property at $z\sim$0 with those at the last simulation snapshot before infall, if they have become satellites between $z\sim$0 and 2 (i.e., $\sim$92\% of the local satellite population). The picture that emerges is significantly different from before, with variations in $V/\sigma$ becoming smaller and galaxies preferentially moving horizontally in the diagram. Interestingly, galaxies with small negative $\Delta (SFR$) (i.e., satellites at the beginning of their quenching phase) show marginally positive $\Delta (V/\sigma)$. This is likely because, despite becoming satellites, galaxies keep acquiring additional angular momentum even after infall. 

\section{Discussion and conclusion}
In this work we have quantified the difference in stellar spin parameter and star formation rate between satellites and main-sequence, rotationally-dominated centrals at $z\sim$0 (matched in stellar mass and ellipticity).  
Satellites in the main-sequence and transition region show very similar stellar kinematic properties to star-forming centrals, and only satellites already in the red sequence have spin parameters significantly (i.e., at least a factor of two) lower than those typical of thin star-forming disks. As our control sample of central galaxies at $z\sim$0  includes only galaxies with stellar spin typical of disk-dominated galaxies, the lack of any major decrease in satellite's $V/\sigma$ in the main sequence rules out significant structural transformation before quenching.

If we use the same matching technique presented in Sec.~3 to estimate the variation of $r$-band S\'ersic index ($\Delta(n)$) and stellar mass surface density ($\Delta(\mu_{*})$, where $\mu_{*}=M_{*}/(2\pi r_{e}^{2})$) for main sequence satellite galaxies, we find that both $\Delta(n)$ and $\Delta(\mu_{*})$ change very little ($\sim$0.08 dex on average, with standard deviations $\sim$0.21 dex and 0.18 dex, respectively), in line with what obtained for $\Delta (V/\sigma)$. This is consistent with \cite{bremer18}, who find no difference in the bulge K-band luminosity between late-type blue sequence and green valley galaxies.

In recent years, several works have proposed a scenario in which a rapid increase in the central galaxy density truncates star formation: i.e., galaxies grow their inner core and then quench (a process sometimes referred to as `compaction'; e.g., \citealp{cheung12,fang13,woo15,zolotov15,tacchella16,wang18}). While originally motivated by studies of central galaxies (e.g., \citealp{cheung12,fang13}), compaction has also been suggested as a viable quenching path for satellite galaxies (e.g., \citealp{wang18}). 

Our findings would appear to rule out any forms of `compaction' that significantly reduces the stellar spin (or increases the average stellar surface density) of main sequence satellites within one effective radius (with respect to star-forming, rotationally-dominated centrals). Given the limited spatial resolution of SAMI observations, we cannot exclude changes in the central kiloparsec of satellite galaxies (i.e., where stellar mass surface densities used to quantify compaction are generally estimated). However, if this is the case, `compaction' does not seem to be affecting the global kinematic and/or photometric properties of group galaxies before quenching. 

In other words, it seems unlikely that, after they have become satellites, galaxies grow prominent dispersion-dominated bulges while still on the main sequence. Of course, star-forming satellites could still harbour small photometric and/or kinematic bulge components, but their structural properties are not different from those of star-forming, rotationally-dominated centrals of similar stellar mass. Our interpretation is in line with \cite{tacchella17} and \cite{abramson18} who show that `compaction' may not be needed to explain the properties of the local passive population.

While SAMI data alone allow us to determine what happens to $z\sim$0 star-forming satellites at the start of their quenching phase, cosmological simulations are invaluable to properly reconstruct the evolution of passive satellites (i.e., galaxies with $\Delta(SFR)<1$ dex). By comparing our findings with predictions from the EAGLE hydrodynamical simulation, we demonstrate that the difference in spin parameter between satellites and centrals must be interpreted as just an upper-limit of the {\it true} structural transformation experienced by satellites after infall. 

\begin{figure}
\centering
\includegraphics[width=8.5cm]{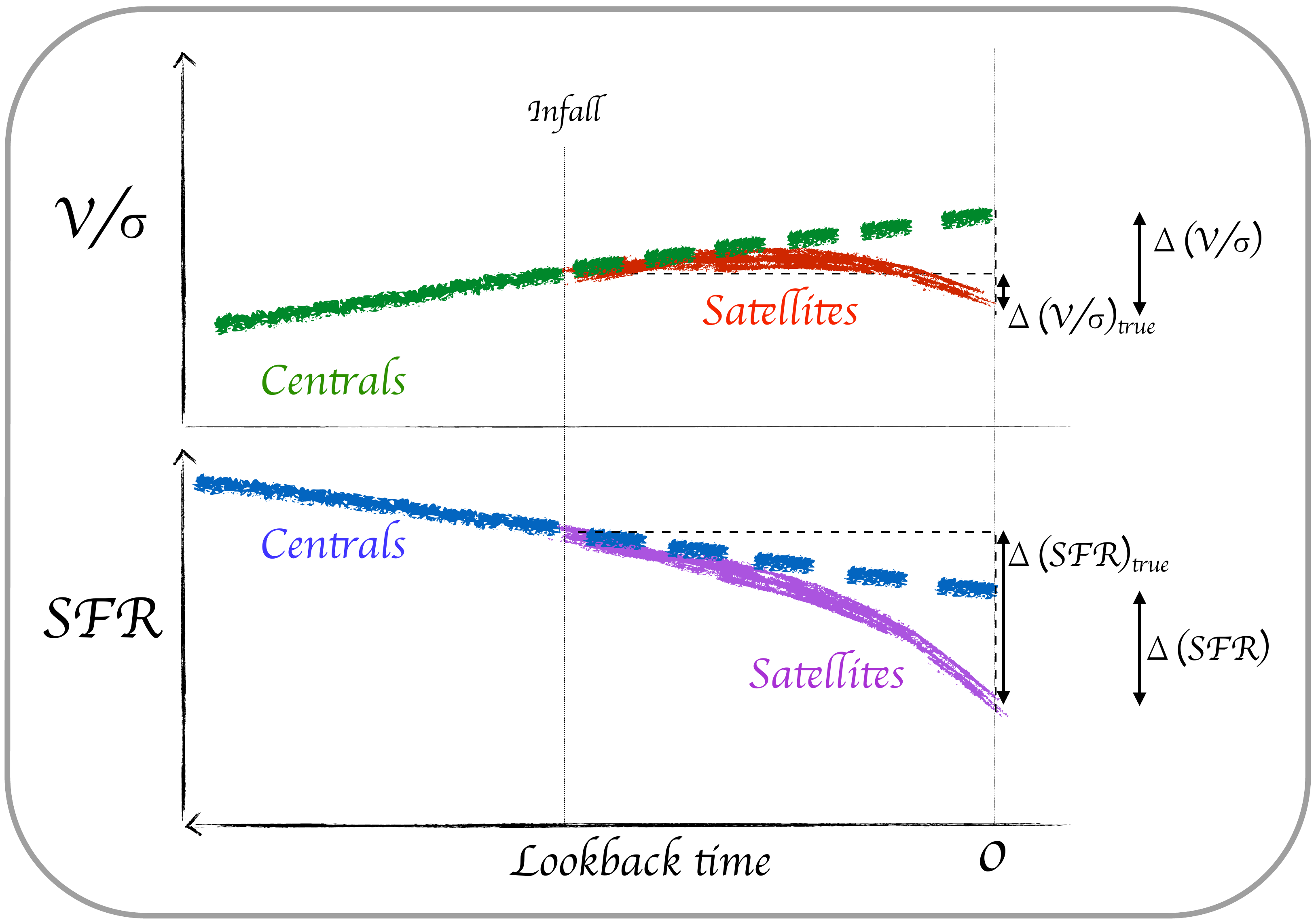}
\caption{A cartoon summarising the evolutionary scenario emerging from this work and the potential effect of {\it progenitor bias}. The top panel shows the increase of $V/\sigma$ with decreasing lookback time/redshift for galaxies while being star-forming centrals (solid green line), and the change in $V/\sigma$ once they become satellites (red line). The green dashed line shows the expected evolution of $V/\sigma$ in case the galaxy would have remained a star-forming central until $z\sim$0. The true $\Delta (V/\sigma)$ and the value obtained via our matching technique are shown by the black vertical arrows. The bottom panel shows the case of $SFR$, with the changes for centrals and satellites highlighted by the blue and pink lines, respectively. In this case, the observed $\Delta (SFR)$ at $z\sim$0 is always smaller than the real value.}
\label{caarton}
\end{figure}

Indeed, at least within the framework of EAGLE, the difference in spin between central and satellites at $z\sim$0 ($\Delta (V/\sigma$)) is always larger than the actual loss experienced by satellites since infall ($\Delta (V/\sigma)_{true}$). This is because most of the observed difference at $z\sim$0 is due to the star-forming central population acquiring additional angular momentum in the last few billion years, rather than satellites losing it via environmental effects during the quenching phase. This is summarised in the cartoon presented in Fig.~\ref{caarton}, which illustrates why Figs.~\ref{deltaeagle} and \ref{deltareal} are so different. 

From theoretical models of structure formation (e.g., \citealp{white84,mo98}), hydrodynamical simulations (e.g., \citealp{pedrosa15,lagos17}), as well as recent observations (e.g., \citealp{simons17,swinbank17}), we see that star-forming central disk galaxies gradually increase their spin with time (solid green line in the top panel), due to the continuing accretion of gas which, on average, is expected to bring high specific angular momentum (e.g., \citealp{catelan96,elbadry18}). 
The typical increase expected in the stellar spin parameter from z$\sim$1 to 0 is $\sim$0.3 dex in our stellar mass range \citep{lagos17}, consistent with the 
observed decrease in gas velocity dispersion \citep{wisnioski15,simons17} and increase in gas specific angular momentum \citep{swinbank17}.  
After infall, the spin of satellite galaxies either remains constant or slightly decreases (solid red line) whereas centrals keep acquiring angular momentum (dashed green line). 
Thus, the difference observed at $z\sim$0 between centrals and satellites is always {\it larger} than the real change in $V/\sigma$ experienced by satellite galaxies.  
The situation is opposite in the case of the $SFR$. On average, a galaxy's star formation activity is decreasing over time (solid blue line, e.g., \citealp{madau96}). Thus, when centrals become satellites the effect of environment is simply to accelerate this decrease. As such, the difference in $SFR$ observed at $z\sim$0 is always {\it lower} than the decrease experienced by satellites since infall.  
In EAGLE we know the properties at infall, so we can relate $z\sim$0 satellites to their progenitors. In the observations, we are forced to compare satellites to $z\sim$0 central, missing the changes that centrals themselves have experienced since the time of infall of satellites into their halos.  

Our results demonstrate that the first and most important phase in the transformation of satellites is quenching, i.e., a significant reduction in their star formation activity. Changes in stellar kinematic properties (i.e., structure) - if any - become evident at a later stage and are on average minor, such that satellites remain rotationally-dominated. This is consistent with a scenario in which multiple physical processes - acting on different time-scales - may play a significant role in altering the evolutionary history of galaxies in groups and clusters. Indeed, while many physical processes (e.g., ram-pressure, tidal stripping) are able to start actively removing the gas reservoir of galaxies and initiate the quenching phase soon after infall, it can take a significantly longer time for low-speed gravitational interactions and/or minor mergers to change the kinematic properties of satellites. However, detailed analysis and modeling of objects occupying different regions in the $\Delta (SFR)$ vs. $\Delta (V/\sigma)$ is required to properly isolate the physical processes acting on satellite galaxies. Moreover, it is important to remember that our results are valid for galaxies with stellar masses greater than 10$^{10}$ M$_{\odot}$ and cannot be blindly extrapolated to lower stellar masses. 

Thanks to the way $\Delta (SFR)$ and $\Delta (V/\sigma)$ are quantified for both SAMI and EAGLE galaxies, they automatically incorporate the effect of any pre-processing on galaxy transformation: i.e., environmental effects experienced by galaxies while satellites in a halo different from the one occupied at $z\sim$0. Thus, our results also suggest that in current numerical simulations pre-processing has a limited effect on the structural properties of satellite galaxies, contrary to what is commonly assumed (e.g., \citealp{zabludoff98,big}). 


\section*{Acknowledgments}
We thank the referee for useful comments and suggestions that improved the clarity of this manuscript.

LC is the recipient of an Australian Research Council Future Fellowship (FT180100066) funded by the Australian Government.
This research was conducted by the Australian Research Council Centre of Excellence for All Sky Astrophysics in 3 Dimensions (ASTRO 3D), through project number CE170100013.
JvdS is funded under Bland-Hawthorn's ARC Laureate Fellowship (FL140100278). 
CL has received funding from a Discovery Early Career Researcher Award (DE150100618) and the MERAC Foundation for a Postdoctoral Research Award. 
SB acknowledges the funding support from the Australian Research Council through a Future Fellowship (FT140101166). M.S.O. acknowledges the funding support from the Australian Research Council through a Future Fellowship (FT140100255). 

The SAMI Galaxy Survey is based on observations made at the Anglo-Australian Telescope. The Sydney-AAO Multi-object Integral field spectrograph (SAMI) was developed jointly by the University of Sydney and the Australian Astronomical Observatory. The SAMI input catalogue is based on data taken from the Sloan Digital Sky Survey, the GAMA Survey and the VST ATLAS Survey. The SAMI Galaxy Survey is supported by the Australian Research Council Centre of Excellence for All Sky Astrophysics in 3 Dimensions (ASTRO 3D), through project number CE170100013, the Australian Research Council Centre of Excellence for All-sky Astrophysics (CAASTRO), through project number CE110001020, and other participating institutions. The SAMI Galaxy Survey website is \url{http://sami-survey.org/}

GAMA is a joint European-Australasian project based around a spectroscopic campaign using the Anglo-Australian Telescope. The GAMA input catalogue is based on data taken from the Sloan Digital Sky Survey and the UKIRT Infrared Deep Sky Survey. Complementary imaging of the GAMA regions is being obtained by a number of independent survey programs including GALEX MIS, VST KiDS, VISTA VIKING, WISE, Herschel-ATLAS, GMRT and ASKAP providing UV to radio coverage. GAMA is funded by the STFC (UK), the ARC (Australia), the AAO, and the participating institutions. The GAMA website is \url{http://www.gama-survey.org/}. 

Part of this work was performed on the gSTAR national facility at Swinburne University of Technology. gSTAR is funded by Swinburne and the Australian Government’s Education Investment Fund.

We acknowledge the Virgo Consortium for making their simulation data available. The EAGLE simulations were performed using the DiRAC-2 facility at Durham, managed by the ICC, and the PRACE facility Curie based in France at TGCC, CEA, Bruyeres-le-Chatel.

\bibliography{main}

\begin{thebibliography}{}
\makeatletter
\relax
\def\mn@urlcharsother{\let\do\@makeother \do\$\do\&\do\#\do\^\do\_\do\%\do\~}
\def\mn@doi{\begingroup\mn@urlcharsother \@ifnextchar [ {\mn@doi@}
  {\mn@doi@[]}}
\def\mn@doi@[#1]#2{\def\@tempa{#1}\ifx\@tempa\@empty \href
  {http://dx.doi.org/#2} {doi:#2}\else \href {http://dx.doi.org/#2} {#1}\fi
  \endgroup}
\def\mn@eprint#1#2{\mn@eprint@#1:#2::\@nil}
\def\mn@eprint@arXiv#1{\href {http://arxiv.org/abs/#1} {{\tt arXiv:#1}}}
\def\mn@eprint@dblp#1{\href {http://dblp.uni-trier.de/rec/bibtex/#1.xml}
  {dblp:#1}}
\def\mn@eprint@#1:#2:#3:#4\@nil{\def\@tempa {#1}\def\@tempb {#2}\def\@tempc
  {#3}\ifx \@tempc \@empty \let \@tempc \@tempb \let \@tempb \@tempa \fi \ifx
  \@tempb \@empty \def\@tempb {arXiv}\fi \@ifundefined
  {mn@eprint@\@tempb}{\@tempb:\@tempc}{\expandafter \expandafter \csname
  mn@eprint@\@tempb\endcsname \expandafter{\@tempc}}}

\bibitem[\protect\citeauthoryear{{Abramson} \& {Morishita}}{{Abramson} \&
  {Morishita}}{2018}]{abramson18}
{Abramson} L.~E.,  {Morishita} T.,  2018, \mn@doi [\apj]
  {10.3847/1538-4357/aab61b}, \href
  {https://ui.adsabs.harvard.edu/#abs/2018ApJ...858...40A} {858, 40}

\bibitem[\protect\citeauthoryear{{Bamford} et~al.,}{{Bamford}
  et~al.}{2009}]{bamford09}
{Bamford} S.~P.,  et~al., 2009, \mn@doi [\mnras]
  {10.1111/j.1365-2966.2008.14252.x}, \href
  {http://adsabs.harvard.edu/abs/2009MNRAS.393.1324B} {393, 1324}

\bibitem[\protect\citeauthoryear{{Bland-Hawthorn} et~al.,}{{Bland-Hawthorn}
  et~al.}{2011}]{bland11}
{Bland-Hawthorn} J.,  et~al., 2011, \mn@doi [Optics Express]
  {10.1364/OE.19.002649}, \href
  {http://adsabs.harvard.edu/abs/2011OExpr..19.2649B} {19, 2649}

\bibitem[\protect\citeauthoryear{{Blanton} \& {Berlind}}{{Blanton} \&
  {Berlind}}{2007}]{blanton07}
{Blanton} M.~R.,  {Berlind} A.~A.,  2007, \mn@doi [\apj] {10.1086/512478},
  \href {https://ui.adsabs.harvard.edu/#abs/2007ApJ...664..791B} {664, 791}

\bibitem[\protect\citeauthoryear{{Blanton}, {Eisenstein}, {Hogg}, {Schlegel}
  \& {Brinkmann}}{{Blanton} et~al.}{2005}]{blanton05}
{Blanton} M.~R.,  {Eisenstein} D.,  {Hogg} D.~W.,  {Schlegel} D.~J.,
  {Brinkmann} J.,  2005, \mn@doi [\apj] {10.1086/422897}, \href
  {http://adsabs.harvard.edu/abs/2005ApJ...629..143B} {629, 143}

\bibitem[\protect\citeauthoryear{{Boselli} \& {Gavazzi}}{{Boselli} \&
  {Gavazzi}}{2006}]{review}
{Boselli} A.,  {Gavazzi} G.,  2006, \mn@doi [\pasp] {10.1086/500691}, \href
  {http://adsabs.harvard.edu/cgi-bin/nph-bib_query?bibcode=2006PASP..118..517B&db_key=AST}
  {118, 517}

\bibitem[\protect\citeauthoryear{{Boselli}, {Boissier}, {Cortese}  \&
  {Gavazzi}}{{Boselli} et~al.}{2008}]{dEale}
{Boselli} A.,  {Boissier} S.,  {Cortese} L.,   {Gavazzi} G.,  2008, \mn@doi
  [\apj] {10.1086/525513}, \href
  {http://adsabs.harvard.edu/abs/2008ApJ...674..742B} {674, 742}

\bibitem[\protect\citeauthoryear{{Bremer} et~al.,}{{Bremer}
  et~al.}{2018}]{bremer18}
{Bremer} M.~N.,  et~al., 2018, \mn@doi [\mnras] {10.1093/mnras/sty124}, \href
  {https://ui.adsabs.harvard.edu/#abs/2018MNRAS.476...12B} {476, 12}

\bibitem[\protect\citeauthoryear{{Brown} et~al.,}{{Brown}
  et~al.}{2017}]{brown17}
{Brown} T.,  et~al., 2017, \mn@doi [\mnras] {10.1093/mnras/stw2991}, \href
  {http://adsabs.harvard.edu/abs/2017MNRAS.466.1275B} {466, 1275}

\bibitem[\protect\citeauthoryear{{Bryant}, {Bland-Hawthorn}, {Fogarty},
  {Lawrence}  \& {Croom}}{{Bryant} et~al.}{2014}]{hexa14}
{Bryant} J.~J.,  {Bland-Hawthorn} J.,  {Fogarty} L.~M.~R.,  {Lawrence} J.~S.,
  {Croom} S.~M.,  2014, \mn@doi [\mnras] {10.1093/mnras/stt2254}, \href
  {http://adsabs.harvard.edu/abs/2014MNRAS.438..869B} {438, 869}

\bibitem[\protect\citeauthoryear{{Bryant} et~al.,}{{Bryant}
  et~al.}{2015}]{bryant15}
{Bryant} J.~J.,  et~al., 2015, \mn@doi [\mnras] {10.1093/mnras/stu2635}, \href
  {http://adsabs.harvard.edu/abs/2015MNRAS.447.2857B} {447, 2857}

\bibitem[\protect\citeauthoryear{{Bundy} et~al.,}{{Bundy}
  et~al.}{2010}]{bundy10}
{Bundy} K.,  et~al., 2010, \mn@doi [\apj] {10.1088/0004-637X/719/2/1969}, \href
  {http://adsabs.harvard.edu/abs/2010ApJ...719.1969B} {719, 1969}

\bibitem[\protect\citeauthoryear{{Cappellari}}{{Cappellari}}{2008}]{cappellari08}
{Cappellari} M.,  2008, \mn@doi [\mnras] {10.1111/j.1365-2966.2008.13754.x},
  \href {https://ui.adsabs.harvard.edu/#abs/2008MNRAS.390...71C} {390, 71}

\bibitem[\protect\citeauthoryear{{Cappellari}}{{Cappellari}}{2013}]{cappellari13}
{Cappellari} M.,  2013, \mn@doi [\apj] {10.1088/2041-8205/778/1/L2}, \href
  {https://ui.adsabs.harvard.edu/#abs/2013ApJ...778L...2C} {778, L2}

\bibitem[\protect\citeauthoryear{{Cappellari}}{{Cappellari}}{2016}]{cappellari16}
{Cappellari} M.,  2016, \mn@doi [Annual Review of Astronomy and Astrophysics]
  {10.1146/annurev-astro-082214-122432}, \href
  {https://ui.adsabs.harvard.edu/#abs/2016ARA&A..54..597C} {54, 597}

\bibitem[\protect\citeauthoryear{{Cappellari} \& {Emsellem}}{{Cappellari} \&
  {Emsellem}}{2004}]{cappellari2004}
{Cappellari} M.,  {Emsellem} E.,  2004, \mn@doi [\pasp] {10.1086/381875}, \href
  {http://adsabs.harvard.edu/abs/2004PASP..116..138C} {116, 138}

\bibitem[\protect\citeauthoryear{{Cappellari} et~al.,}{{Cappellari}
  et~al.}{2007}]{cappellari07}
{Cappellari} M.,  et~al., 2007, \mn@doi [\mnras]
  {10.1111/j.1365-2966.2007.11963.x}, \href
  {http://adsabs.harvard.edu/abs/2007MNRAS.379..418C} {379, 418}

\bibitem[\protect\citeauthoryear{{Catelan} \& {Theuns}}{{Catelan} \&
  {Theuns}}{1996}]{catelan96}
{Catelan} P.,  {Theuns} T.,  1996, \mn@doi [\mnras] {10.1093/mnras/282.2.436},
  \href {https://ui.adsabs.harvard.edu/#abs/1996MNRAS.282..436C} {282, 436}

\bibitem[\protect\citeauthoryear{{Cheung} et~al.,}{{Cheung}
  et~al.}{2012}]{cheung12}
{Cheung} E.,  et~al., 2012, \mn@doi [\apj] {10.1088/0004-637X/760/2/131}, \href
  {https://ui.adsabs.harvard.edu/#abs/2012ApJ...760..131C} {760, 131}

\bibitem[\protect\citeauthoryear{{Christlein} \& {Zabludoff}}{{Christlein} \&
  {Zabludoff}}{2004}]{christlein04}
{Christlein} D.,  {Zabludoff} A.~I.,  2004, \mn@doi [\apj] {10.1086/424909},
  \href
  {http://adsabs.harvard.edu/cgi-bin/nph-bib_query?bibcode=2004ApJ...616..192C&db_key=AST}
  {616, 192}

\bibitem[\protect\citeauthoryear{{Correa}, {Schaye}, {Clauwens}, {Bower},
  {Crain}, {Schaller}, {Theuns}  \& {Thob}}{{Correa} et~al.}{2017}]{correa17}
{Correa} C.~A.,  {Schaye} J.,  {Clauwens} B.,  {Bower} R.~G.,  {Crain} R.~A.,
  {Schaller} M.,  {Theuns} T.,   {Thob} A. C.~R.,  2017, \mn@doi [\mnras]
  {10.1093/mnrasl/slx133}, \href
  {https://ui.adsabs.harvard.edu/#abs/2017MNRAS.472L..45C} {472, L45}

\bibitem[\protect\citeauthoryear{{Cortese} \& {Hughes}}{{Cortese} \&
  {Hughes}}{2009}]{cortese09}
{Cortese} L.,  {Hughes} T.~M.,  2009, \mn@doi [\mnras]
  {10.1111/j.1365-2966.2009.15548.x}, \href
  {http://adsabs.harvard.edu/abs/2009MNRAS.400.1225C} {400, 1225}

\bibitem[\protect\citeauthoryear{{Cortese}, {Gavazzi}, {Boselli}, {Franzetti},
  {Kennicutt}, {O'Neil}  \& {Sakai}}{{Cortese} et~al.}{2006}]{big}
{Cortese} L.,  {Gavazzi} G.,  {Boselli} A.,  {Franzetti} P.,  {Kennicutt}
  R.~C.,  {O'Neil} K.,   {Sakai} S.,  2006, \mn@doi [\aap]
  {10.1051/0004-6361:20064873}, \href
  {http://adsabs.harvard.edu/cgi-bin/nph-bib_query?bibcode=2006A%26A...453..847C&db_key=AST}
  {453, 847}

\bibitem[\protect\citeauthoryear{{Cortese}, {Catinella}, {Boissier}, {Boselli}
  \& {Heinis}}{{Cortese} et~al.}{2011}]{cortese11}
{Cortese} L.,  {Catinella} B.,  {Boissier} S.,  {Boselli} A.,   {Heinis} S.,
  2011, \mn@doi [\mnras] {10.1111/j.1365-2966.2011.18822.x}, \href
  {http://adsabs.harvard.edu/abs/2011MNRAS.415.1797C} {415, 1797}

\bibitem[\protect\citeauthoryear{{Cortese} et~al.,}{{Cortese}
  et~al.}{2016a}]{cortese16b}
{Cortese} L.,  et~al., 2016a, \mn@doi [\mnras] {10.1093/mnras/stw801}, \href
  {http://adsabs.harvard.edu/abs/2016MNRAS.459.3574C} {459, 3574}

\bibitem[\protect\citeauthoryear{{Cortese} et~al.,}{{Cortese}
  et~al.}{2016b}]{cortese16}
{Cortese} L.,  et~al., 2016b, \mn@doi [\mnras] {10.1093/mnras/stw1891}, \href
  {http://adsabs.harvard.edu/abs/2016MNRAS.463..170C} {463, 170}

\bibitem[\protect\citeauthoryear{{Crain} et~al.,}{{Crain}
  et~al.}{2015}]{crain15}
{Crain} R.~A.,  et~al., 2015, \mn@doi [\mnras] {10.1093/mnras/stv725}, \href
  {https://ui.adsabs.harvard.edu/#abs/2015MNRAS.450.1937C} {450, 1937}

\bibitem[\protect\citeauthoryear{{Croom} et~al.,}{{Croom}
  et~al.}{2012}]{croom12}
{Croom} S.~M.,  et~al., 2012, \mn@doi [\mnras]
  {10.1111/j.1365-2966.2011.20365.x}, \href
  {http://adsabs.harvard.edu/abs/2012MNRAS.421..872C} {421, 872}

\bibitem[\protect\citeauthoryear{{Davies} et~al.,}{{Davies}
  et~al.}{2016}]{davies16}
{Davies} L.~J.~M.,  et~al., 2016, \mn@doi [\mnras] {10.1093/mnras/stw1342},
  \href {http://adsabs.harvard.edu/abs/2016MNRAS.461..458D} {461, 458}

\bibitem[\protect\citeauthoryear{{De Lucia}, {Weinmann}, {Poggianti},
  {Arag{\'o}n-Salamanca}  \& {Zaritsky}}{{De Lucia} et~al.}{2012}]{delucia12}
{De Lucia} G.,  {Weinmann} S.,  {Poggianti} B.~M.,  {Arag{\'o}n-Salamanca} A.,
   {Zaritsky} D.,  2012, \mn@doi [\mnras] {10.1111/j.1365-2966.2012.20983.x},
  \href {http://adsabs.harvard.edu/abs/2012MNRAS.423.1277D} {423, 1277}

\bibitem[\protect\citeauthoryear{{Dressler}}{{Dressler}}{1980}]{dressler80}
{Dressler} A.,  1980, \apj, \href
  {http://adsabs.harvard.edu/cgi-bin/nph-bib_query?bibcode=1980ApJ...236..351D&db_key=AST}
  {236, 351}

\bibitem[\protect\citeauthoryear{{Driver} et~al.,}{{Driver}
  et~al.}{2011}]{gama}
{Driver} S.~P.,  et~al., 2011, \mn@doi [\mnras]
  {10.1111/j.1365-2966.2010.18188.x}, \href
  {http://adsabs.harvard.edu/abs/2011MNRAS.413..971D} {413, 971}

\bibitem[\protect\citeauthoryear{{Driver} et~al.,}{{Driver}
  et~al.}{2016}]{driver16}
{Driver} S.~P.,  et~al., 2016, \mn@doi [\mnras] {10.1093/mnras/stv2505}, \href
  {http://adsabs.harvard.edu/abs/2016MNRAS.455.3911D} {455, 3911}

\bibitem[\protect\citeauthoryear{{El-Badry} et~al.,}{{El-Badry}
  et~al.}{2018}]{elbadry18}
{El-Badry} K.,  et~al., 2018, \mn@doi [\mnras] {10.1093/mnras/stx2482}, \href
  {https://ui.adsabs.harvard.edu/#abs/2018MNRAS.473.1930E} {473, 1930}

\bibitem[\protect\citeauthoryear{{Ellison}, {Fertig}, {Rosenberg}, {Nair},
  {Simard}, {Torrey}  \& {Patton}}{{Ellison} et~al.}{2015}]{ellison15}
{Ellison} S.~L.,  {Fertig} D.,  {Rosenberg} J.~L.,  {Nair} P.,  {Simard} L.,
  {Torrey} P.,   {Patton} D.~R.,  2015, \mn@doi [\mnras]
  {10.1093/mnras/stu2744}, \href
  {http://adsabs.harvard.edu/abs/2015MNRAS.448..221E} {448, 221}

\bibitem[\protect\citeauthoryear{{Ellison}, {Catinella}  \&
  {Cortese}}{{Ellison} et~al.}{2018}]{ellison18}
{Ellison} S.~L.,  {Catinella} B.,   {Cortese} L.,  2018, \mn@doi [\mnras]
  {10.1093/mnras/sty1247}, \href
  {http://adsabs.harvard.edu/abs/2018MNRAS.478.3447E} {478, 3447}

\bibitem[\protect\citeauthoryear{{Emsellem} et~al.,}{{Emsellem}
  et~al.}{2007}]{emsellem07}
{Emsellem} E.,  et~al., 2007, \mn@doi [\mnras]
  {10.1111/j.1365-2966.2007.11752.x}, \href
  {http://adsabs.harvard.edu/abs/2007MNRAS.379..401E} {379, 401}

\bibitem[\protect\citeauthoryear{{Emsellem} et~al.,}{{Emsellem}
  et~al.}{2011}]{emsellem11}
{Emsellem} E.,  et~al., 2011, \mn@doi [\mnras]
  {10.1111/j.1365-2966.2011.18496.x}, \href
  {http://adsabs.harvard.edu/abs/2011MNRAS.414..888E} {414, 888}

\bibitem[\protect\citeauthoryear{{Fang}, {Faber}, {Koo}  \& {Dekel}}{{Fang}
  et~al.}{2013}]{fang13}
{Fang} J.~J.,  {Faber} S.~M.,  {Koo} D.~C.,   {Dekel} A.,  2013, \mn@doi [\apj]
  {10.1088/0004-637X/776/1/63}, \href
  {https://ui.adsabs.harvard.edu/#abs/2013ApJ...776...63F} {776, 63}

\bibitem[\protect\citeauthoryear{{Fossati} et~al.,}{{Fossati}
  et~al.}{2015}]{fossati15}
{Fossati} M.,  et~al., 2015, \mn@doi [\mnras] {10.1093/mnras/stu2255}, \href
  {http://adsabs.harvard.edu/abs/2015MNRAS.446.2582F} {446, 2582}

\bibitem[\protect\citeauthoryear{{Foster} et~al.,}{{Foster}
  et~al.}{2017}]{foster17}
{Foster} C.,  et~al., 2017, \mn@doi [\mnras] {10.1093/mnras/stx1869}, \href
  {http://adsabs.harvard.edu/abs/2017MNRAS.472..966F} {472, 966}

\bibitem[\protect\citeauthoryear{{Furlong} et~al.,}{{Furlong}
  et~al.}{2015}]{furlong15}
{Furlong} M.,  et~al., 2015, \mn@doi [\mnras] {10.1093/mnras/stv852}, \href
  {https://ui.adsabs.harvard.edu/#abs/2015MNRAS.450.4486F} {450, 4486}

\bibitem[\protect\citeauthoryear{{Gavazzi}, {Boselli}, {Cortese}, {Arosio},
  {Gallazzi}, {Pedotti}  \& {Carrasco}}{{Gavazzi} et~al.}{2006}]{ha06}
{Gavazzi} G.,  {Boselli} A.,  {Cortese} L.,  {Arosio} I.,  {Gallazzi} A.,
  {Pedotti} P.,   {Carrasco} L.,  2006, \mn@doi [\aap]
  {10.1051/0004-6361:20053843}, \href
  {http://adsabs.harvard.edu/cgi-bin/nph-bib_query?bibcode=2006A%26A...446..839G&db_key=AST}
  {446, 839}

\bibitem[\protect\citeauthoryear{{George}, {Ma}, {Bundy}, {Leauthaud},
  {Tinker}, {Wechsler}, {Finoguenov}  \& {Vulcani}}{{George}
  et~al.}{2013}]{george13}
{George} M.~R.,  {Ma} C.-P.,  {Bundy} K.,  {Leauthaud} A.,  {Tinker} J.,
  {Wechsler} R.~H.,  {Finoguenov} A.,   {Vulcani} B.,  2013, \mn@doi [\apj]
  {10.1088/0004-637X/770/2/113}, \href
  {http://adsabs.harvard.edu/abs/2013ApJ...770..113G} {770, 113}

\bibitem[\protect\citeauthoryear{{Giovanelli} \& {Haynes}}{{Giovanelli} \&
  {Haynes}}{1985}]{giova85}
{Giovanelli} R.,  {Haynes} M.~P.,  1985, \mn@doi [\apj] {10.1086/163170}, \href
  {http://adsabs.harvard.edu/cgi-bin/nph-bib_query?bibcode=1985ApJ...292..404G&db_key=AST}
  {292, 404}

\bibitem[\protect\citeauthoryear{{Giovanelli}, {Haynes}, {Salzer}, {Wegner},
  {da Costa}  \& {Freudling}}{{Giovanelli} et~al.}{1994}]{giovanelli94}
{Giovanelli} R.,  {Haynes} M.~P.,  {Salzer} J.~J.,  {Wegner} G.,  {da Costa}
  L.~N.,   {Freudling} W.,  1994, \mn@doi [\aj] {10.1086/117014}, \href
  {http://adsabs.harvard.edu/abs/1994AJ....107.2036G} {107, 2036}

\bibitem[\protect\citeauthoryear{{G{\'o}mez} et~al.,}{{G{\'o}mez}
  et~al.}{2003}]{gomez03}
{G{\'o}mez} P.~L.,  et~al., 2003, \mn@doi [\apj] {10.1086/345593}, \href
  {http://adsabs.harvard.edu/cgi-bin/nph-bib_query?bibcode=2003ApJ...584..210G&db_key=AST}
  {584, 210}

\bibitem[\protect\citeauthoryear{{Graham} et~al.,}{{Graham}
  et~al.}{2018}]{graham18}
{Graham} M.~T.,  et~al., 2018, \mn@doi [\mnras] {10.1093/mnras/sty504}, \href
  {https://ui.adsabs.harvard.edu/#abs/2018MNRAS.477.4711G} {477, 4711}

\bibitem[\protect\citeauthoryear{{Han}, {Smith}, {Choi}, {Cortese},
  {Catinella}, {Contini}  \& {Yi}}{{Han} et~al.}{2018}]{han18}
{Han} S.,  {Smith} R.,  {Choi} H.,  {Cortese} L.,  {Catinella} B.,  {Contini}
  E.,   {Yi} S.~K.,  2018, \mn@doi [\apj] {10.3847/1538-4357/aadfe2}, \href
  {http://adsabs.harvard.edu/abs/2018ApJ...866...78H} {866, 78}

\bibitem[\protect\citeauthoryear{{Harborne}, {Power}, {Robotham}, {Cortese}  \&
  {Taranu}}{{Harborne} et~al.}{2019}]{harborne19}
{Harborne} K.~E.,  {Power} C.,  {Robotham} A.~S.~G.,  {Cortese} L.,   {Taranu}
  D.~S.,  2019, \mn@doi [\mnras] {10.1093/mnras/sty3120}, \href
  {http://adsabs.harvard.edu/abs/2019MNRAS.483..249H} {483, 249}

\bibitem[\protect\citeauthoryear{{Haynes} \& {Giovanelli}}{{Haynes} \&
  {Giovanelli}}{1984}]{haynes}
{Haynes} M.~P.,  {Giovanelli} R.,  1984, \mn@doi [\aj] {10.1086/113573}, \href
  {http://adsabs.harvard.edu/cgi-bin/nph-bib_query?bibcode=1984AJ.....89..758H&db_key=AST}
  {89, 758}

\bibitem[\protect\citeauthoryear{{Hester}}{{Hester}}{2010}]{hester10}
{Hester} J.~A.,  2010, \mn@doi [\apj] {10.1088/0004-637X/720/1/191}, \href
  {https://ui.adsabs.harvard.edu/#abs/2010ApJ...720..191H} {720, 191}

\bibitem[\protect\citeauthoryear{{Hubble} \& {Humason}}{{Hubble} \&
  {Humason}}{1931}]{HUBH31}
{Hubble} E.,  {Humason} M.~L.,  1931, \apj, \href
  {http://adsabs.harvard.edu/cgi-bin/nph-bib_query?bibcode=1931ApJ....74...43H&db_key=AST}
  {74, 43}

\bibitem[\protect\citeauthoryear{{Kawinwanichakij} et~al.,}{{Kawinwanichakij}
  et~al.}{2017}]{Kawinwanichakij17}
{Kawinwanichakij} L.,  et~al., 2017, \mn@doi [\apj] {10.3847/1538-4357/aa8b75},
  \href {https://ui.adsabs.harvard.edu/#abs/2017ApJ...847..134K} {847, 134}

\bibitem[\protect\citeauthoryear{{Kelvin} et~al.,}{{Kelvin}
  et~al.}{2012}]{kelvin12}
{Kelvin} L.~S.,  et~al., 2012, \mn@doi [\mnras]
  {10.1111/j.1365-2966.2012.20355.x}, \href
  {http://adsabs.harvard.edu/abs/2012MNRAS.421.1007K} {421, 1007}

\bibitem[\protect\citeauthoryear{{Krajnovi{\'c}} et~al.,}{{Krajnovi{\'c}}
  et~al.}{2013}]{krajnovic13}
{Krajnovi{\'c}} D.,  et~al., 2013, \mn@doi [\mnras] {10.1093/mnras/sts315},
  \href {http://adsabs.harvard.edu/abs/2013MNRAS.432.1768K} {432, 1768}

\bibitem[\protect\citeauthoryear{{Lagos}, {Theuns}, {Stevens}, {Cortese},
  {Padilla}, {Davis}, {Contreras}  \& {Croton}}{{Lagos} et~al.}{2017}]{lagos17}
{Lagos} C. d.~P.,  {Theuns} T.,  {Stevens} A. R.~H.,  {Cortese} L.,  {Padilla}
  N.~D.,  {Davis} T.~A.,  {Contreras} S.,   {Croton} D.,  2017, \mn@doi
  [\mnras] {10.1093/mnras/stw2610}, \href
  {https://ui.adsabs.harvard.edu/#abs/2017MNRAS.464.3850L} {464, 3850}

\bibitem[\protect\citeauthoryear{{Lagos}, {Schaye}, {Bah{\'e}}, {Van de Sande},
  {Kay}, {Barnes}, {Davis}  \& {Dalla Vecchia}}{{Lagos} et~al.}{2018}]{lagos18}
{Lagos} C.~d.~P.,  {Schaye} J.,  {Bah{\'e}} Y.,  {Van de Sande} J.,  {Kay}
  S.~T.,  {Barnes} D.,  {Davis} T.~A.,   {Dalla Vecchia} C.,  2018, \mn@doi
  [\mnras] {10.1093/mnras/sty489}, \href
  {http://adsabs.harvard.edu/abs/2018MNRAS.476.4327L} {476, 4327}

\bibitem[\protect\citeauthoryear{{Larson}, {Tinsley}  \& {Caldwell}}{{Larson}
  et~al.}{1980}]{LARS80}
{Larson} R.~B.,  {Tinsley} B.~M.,   {Caldwell} C.~N.,  1980, \apj, \href
  {http://adsabs.harvard.edu/cgi-bin/nph-bib_query?bibcode=1980ApJ...237..692L&db_key=AST}
  {237, 692}

\bibitem[\protect\citeauthoryear{{Lewis} et~al.,}{{Lewis}
  et~al.}{2002}]{lewis02}
{Lewis} I.,  et~al., 2002, \mn@doi [\mnras] {10.1046/j.1365-8711.2002.05558.x},
  \href
  {http://adsabs.harvard.edu/cgi-bin/nph-bib_query?bibcode=2002MNRAS.334..673L&db_key=AST}
  {334, 673}

\bibitem[\protect\citeauthoryear{{Lisker}, {Grebel}  \& {Binggeli}}{{Lisker}
  et~al.}{2006}]{lisker06}
{Lisker} T.,  {Grebel} E.~K.,   {Binggeli} B.,  2006, \mn@doi [\aj]
  {10.1086/505045}, \href
  {https://ui.adsabs.harvard.edu/#abs/2006AJ....132..497L} {132, 497}

\bibitem[\protect\citeauthoryear{{Madau}, {Ferguson}, {Dickinson},
  {Giavalisco}, {Steidel}  \& {Fruchter}}{{Madau} et~al.}{1996}]{madau96}
{Madau} P.,  {Ferguson} H.~C.,  {Dickinson} M.~E.,  {Giavalisco} M.,  {Steidel}
  C.~C.,   {Fruchter} A.,  1996, \mnras, \href
  {http://adsabs.harvard.edu/abs/1996MNRAS.283.1388M} {283, 1388}

\bibitem[\protect\citeauthoryear{{McAlpine} et~al.,}{{McAlpine}
  et~al.}{2016}]{mcalpine16}
{McAlpine} S.,  et~al., 2016, \mn@doi [Astronomy and Computing]
  {10.1016/j.ascom.2016.02.004}, \href
  {https://ui.adsabs.harvard.edu/#abs/2016A&C....15...72M} {15, 72}

\bibitem[\protect\citeauthoryear{{Mei} et~al.,}{{Mei} et~al.}{2007}]{mei07}
{Mei} S.,  et~al., 2007, \mn@doi [\apj] {10.1086/509598}, \href
  {http://adsabs.harvard.edu/abs/2007ApJ...655..144M} {655, 144}

\bibitem[\protect\citeauthoryear{{Mo}, {Mao}  \& {White}}{{Mo}
  et~al.}{1998}]{mo98}
{Mo} H.~J.,  {Mao} S.,   {White} S.~D.~M.,  1998, \mn@doi [\mnras]
  {10.1046/j.1365-8711.1998.01227.x}, \href
  {http://adsabs.harvard.edu/abs/1998MNRAS.295..319M} {295, 319}

\bibitem[\protect\citeauthoryear{{Moss} \& {Whittle}}{{Moss} \&
  {Whittle}}{2000}]{moss00}
{Moss} C.,  {Whittle} M.,  2000, \mnras, \href
  {http://adsabs.harvard.edu/abs/2000MNRAS.317..667M} {317, 667}

\bibitem[\protect\citeauthoryear{{Muldrew} et~al.,}{{Muldrew}
  et~al.}{2012}]{muldrew12}
{Muldrew} S.~I.,  et~al., 2012, \mn@doi [\mnras]
  {10.1111/j.1365-2966.2011.19922.x}, \href
  {http://adsabs.harvard.edu/abs/2012MNRAS.419.2670M} {419, 2670}

\bibitem[\protect\citeauthoryear{{Oman} \& {Hudson}}{{Oman} \&
  {Hudson}}{2016}]{oman16}
{Oman} K.~A.,  {Hudson} M.~J.,  2016, \mn@doi [\mnras] {10.1093/mnras/stw2195},
  \href {https://ui.adsabs.harvard.edu/#abs/2016MNRAS.463.3083O} {463, 3083}

\bibitem[\protect\citeauthoryear{{Omand}, {Balogh}  \& {Poggianti}}{{Omand}
  et~al.}{2014}]{omand14}
{Omand} C.~M.~B.,  {Balogh} M.~L.,   {Poggianti} B.~M.,  2014, \mn@doi [\mnras]
  {10.1093/mnras/stu331}, \href
  {http://adsabs.harvard.edu/abs/2014MNRAS.440..843O} {440, 843}

\bibitem[\protect\citeauthoryear{{Pedrosa} \& {Tissera}}{{Pedrosa} \&
  {Tissera}}{2015}]{pedrosa15}
{Pedrosa} S.~E.,  {Tissera} P.~B.,  2015, \mn@doi [\aap]
  {10.1051/0004-6361/201526440}, \href
  {https://ui.adsabs.harvard.edu/#abs/2015A&A...584A..43P} {584, A43}

\bibitem[\protect\citeauthoryear{{Peng}, {Lilly}, {Renzini}  \&
  {Carollo}}{{Peng} et~al.}{2012}]{peng12}
{Peng} Y.-j.,  {Lilly} S.~J.,  {Renzini} A.,   {Carollo} M.,  2012, \mn@doi
  [\apj] {10.1088/0004-637X/757/1/4}, \href
  {https://ui.adsabs.harvard.edu/#abs/2012ApJ...757....4P} {757, 4}

\bibitem[\protect\citeauthoryear{{Poggianti}, {Smail}, {Dressler}, {Couch},
  {Barger}, {Butcher}, {Ellis}  \& {Oemler}}{{Poggianti}
  et~al.}{1999}]{poggianti99}
{Poggianti} B.~M.,  {Smail} I.,  {Dressler} A.,  {Couch} W.~J.,  {Barger}
  A.~J.,  {Butcher} H.,  {Ellis} R.~S.,   {Oemler} A.~J.,  1999, \mn@doi [\apj]
  {10.1086/307322}, \href {http://adsabs.harvard.edu/abs/1999ApJ...518..576P}
  {518, 576}

\bibitem[\protect\citeauthoryear{{Rizzo}, {Fraternali}  \& {Iorio}}{{Rizzo}
  et~al.}{2018}]{rizzo18}
{Rizzo} F.,  {Fraternali} F.,   {Iorio} G.,  2018, \mn@doi [\mnras]
  {10.1093/mnras/sty347}, \href
  {http://adsabs.harvard.edu/abs/2018MNRAS.476.2137R} {476, 2137}

\bibitem[\protect\citeauthoryear{{Robotham} et~al.,}{{Robotham}
  et~al.}{2011}]{robotham11}
{Robotham} A.~S.~G.,  et~al., 2011, \mn@doi [\mnras]
  {10.1111/j.1365-2966.2011.19217.x}, \href
  {http://adsabs.harvard.edu/abs/2011MNRAS.416.2640R} {416, 2640}

\bibitem[\protect\citeauthoryear{{S{\'a}nchez-Bl{\'a}zquez}
  et~al.,}{{S{\'a}nchez-Bl{\'a}zquez} et~al.}{2006}]{miles}
{S{\'a}nchez-Bl{\'a}zquez} P.,  et~al., 2006, \mn@doi [\mnras]
  {10.1111/j.1365-2966.2006.10699.x}, \href
  {http://adsabs.harvard.edu/abs/2006MNRAS.371..703S} {371, 703}

\bibitem[\protect\citeauthoryear{{Schaye} \& {Dalla Vecchia}}{{Schaye} \&
  {Dalla Vecchia}}{2008}]{schaye08}
{Schaye} J.,  {Dalla Vecchia} C.,  2008, \mn@doi [\mnras]
  {10.1111/j.1365-2966.2007.12639.x}, \href
  {http://adsabs.harvard.edu/abs/2008MNRAS.383.1210S} {383, 1210}

\bibitem[\protect\citeauthoryear{{Schaye} et~al.,}{{Schaye}
  et~al.}{2015}]{eagle15}
{Schaye} J.,  et~al., 2015, \mn@doi [\mnras] {10.1093/mnras/stu2058}, \href
  {http://adsabs.harvard.edu/abs/2015MNRAS.446..521S} {446, 521}

\bibitem[\protect\citeauthoryear{{Scott} et~al.,}{{Scott}
  et~al.}{2018}]{scott2018}
{Scott} N.,  et~al., 2018, \mn@doi [\mnras] {10.1093/mnras/sty2355}, \href
  {http://adsabs.harvard.edu/abs/2018MNRAS.481.2299S} {481, 2299}

\bibitem[\protect\citeauthoryear{{Simons} et~al.,}{{Simons}
  et~al.}{2017}]{simons17}
{Simons} R.~C.,  et~al., 2017, \mn@doi [\apj] {10.3847/1538-4357/aa740c}, \href
  {https://ui.adsabs.harvard.edu/#abs/2017ApJ...843...46S} {843, 46}

\bibitem[\protect\citeauthoryear{{Solanes}, {Manrique},
  {Garc{\'{\i}}a-G{\'o}mez}, {Gonz{\'a}lez-Casado}, {Giovanelli}  \&
  {Haynes}}{{Solanes} et~al.}{2001}]{solanes01}
{Solanes} J.~M.,  {Manrique} A.,  {Garc{\'{\i}}a-G{\'o}mez} C.,
  {Gonz{\'a}lez-Casado} G.,  {Giovanelli} R.,   {Haynes} M.~P.,  2001, \mn@doi
  [\apj] {10.1086/318672}, \href
  {http://adsabs.harvard.edu/abs/2001ApJ...548...97S} {548, 97}

\bibitem[\protect\citeauthoryear{{Swinbank} et~al.,}{{Swinbank}
  et~al.}{2017}]{swinbank17}
{Swinbank} A.~M.,  et~al., 2017, \mn@doi [\mnras] {10.1093/mnras/stx201}, \href
  {https://ui.adsabs.harvard.edu/#abs/2017MNRAS.467.3140S} {467, 3140}

\bibitem[\protect\citeauthoryear{{Tacchella}, {Dekel}, {Carollo}, {Ceverino},
  {DeGraf}, {Lapiner}, {Mandelker}  \& {Primack Joel}}{{Tacchella}
  et~al.}{2016}]{tacchella16}
{Tacchella} S.,  {Dekel} A.,  {Carollo} C.~M.,  {Ceverino} D.,  {DeGraf} C.,
  {Lapiner} S.,  {Mandelker} N.,   {Primack Joel} R.,  2016, \mn@doi [\mnras]
  {10.1093/mnras/stw131}, \href
  {https://ui.adsabs.harvard.edu/#abs/2016MNRAS.457.2790T} {457, 2790}

\bibitem[\protect\citeauthoryear{{Tacchella}, {Carollo}, {Faber}, {Cibinel},
  {Dekel}, {Koo}, {Renzini}  \& {Woo}}{{Tacchella} et~al.}{2017}]{tacchella17}
{Tacchella} S.,  {Carollo} C.~M.,  {Faber} S.~M.,  {Cibinel} A.,  {Dekel} A.,
  {Koo} D.~C.,  {Renzini} A.,   {Woo} J.,  2017, \mn@doi [\apj]
  {10.3847/2041-8213/aa7cfb}, \href
  {https://ui.adsabs.harvard.edu/#abs/2017ApJ...844L...1T} {844, L1}

\bibitem[\protect\citeauthoryear{{Taylor} et~al.,}{{Taylor}
  et~al.}{2011}]{taylor11}
{Taylor} E.~N.,  et~al., 2011, \mn@doi [\mnras]
  {10.1111/j.1365-2966.2011.19536.x}, \href
  {http://adsabs.harvard.edu/abs/2011MNRAS.418.1587T} {418, 1587}

\bibitem[\protect\citeauthoryear{{Tempel}, {Einasto}, {Einasto}, {Saar}  \&
  {Tago}}{{Tempel} et~al.}{2009}]{tempel09}
{Tempel} E.,  {Einasto} J.,  {Einasto} M.,  {Saar} E.,   {Tago} E.,  2009,
  \mn@doi [\aap] {10.1051/0004-6361:200810274}, \href
  {http://adsabs.harvard.edu/abs/2009A%26A...495...37T} {495, 37}

\bibitem[\protect\citeauthoryear{{Toloba} et~al.,}{{Toloba}
  et~al.}{2009}]{toloba09}
{Toloba} E.,  et~al., 2009, \mn@doi [\apj] {10.1088/0004-637X/707/1/L17}, \href
  {https://ui.adsabs.harvard.edu/#abs/2009ApJ...707L..17T} {707, L17}

\bibitem[\protect\citeauthoryear{{Unterborn} \& {Ryden}}{{Unterborn} \&
  {Ryden}}{2008}]{Unterborn08}
{Unterborn} C.~T.,  {Ryden} B.~S.,  2008, \mn@doi [\apj] {10.1086/591898},
  \href {http://adsabs.harvard.edu/abs/2008ApJ...687..976U} {687, 976}

\bibitem[\protect\citeauthoryear{{Wang}, {Kong}  \& {Pan}}{{Wang}
  et~al.}{2018}]{wang18}
{Wang} E.,  {Kong} X.,   {Pan} Z.,  2018, \mn@doi [\apj]
  {10.3847/1538-4357/aadb9e}, \href
  {https://ui.adsabs.harvard.edu/#abs/2018ApJ...865...49W} {865, 49}

\bibitem[\protect\citeauthoryear{{Weinmann}, {Kauffmann}, {van den Bosch},
  {Pasquali}, {McIntosh}, {Mo}, {Yang}  \& {Guo}}{{Weinmann}
  et~al.}{2009}]{weinmann09}
{Weinmann} S.~M.,  {Kauffmann} G.,  {van den Bosch} F.~C.,  {Pasquali} A.,
  {McIntosh} D.~H.,  {Mo} H.,  {Yang} X.,   {Guo} Y.,  2009, \mn@doi [\mnras]
  {10.1111/j.1365-2966.2009.14412.x}, \href
  {https://ui.adsabs.harvard.edu/#abs/2009MNRAS.394.1213W} {394, 1213}

\bibitem[\protect\citeauthoryear{{Weinmann}, {Kauffmann}, {von der Linden}  \&
  {De Lucia}}{{Weinmann} et~al.}{2010}]{weinmann10}
{Weinmann} S.~M.,  {Kauffmann} G.,  {von der Linden} A.,   {De Lucia} G.,
  2010, \mn@doi [\mnras] {10.1111/j.1365-2966.2010.16855.x}, \href
  {https://ui.adsabs.harvard.edu/#abs/2010MNRAS.406.2249W} {406, 2249}

\bibitem[\protect\citeauthoryear{{Wetzel}, {Tinker}  \& {Conroy}}{{Wetzel}
  et~al.}{2012}]{wetzel12}
{Wetzel} A.~R.,  {Tinker} J.~L.,   {Conroy} C.,  2012, \mn@doi [\mnras]
  {10.1111/j.1365-2966.2012.21188.x}, \href
  {https://ui.adsabs.harvard.edu/#abs/2012MNRAS.424..232W} {424, 232}

\bibitem[\protect\citeauthoryear{{Wetzel}, {Tinker}, {Conroy}  \& {van den
  Bosch}}{{Wetzel} et~al.}{2013}]{wetzel13}
{Wetzel} A.~R.,  {Tinker} J.~L.,  {Conroy} C.,   {van den Bosch} F.~C.,  2013,
  \mn@doi [\mnras] {10.1093/mnras/stt469}, \href
  {https://ui.adsabs.harvard.edu/#abs/2013MNRAS.432..336W} {432, 336}

\bibitem[\protect\citeauthoryear{{White}}{{White}}{1984}]{white84}
{White} S.~D.~M.,  1984, \mn@doi [\apj] {10.1086/162573}, \href
  {https://ui.adsabs.harvard.edu/#abs/1984ApJ...286...38W} {286, 38}

\bibitem[\protect\citeauthoryear{{Wilman}, {Zibetti}  \&
  {Budav{\'a}ri}}{{Wilman} et~al.}{2010}]{willman10}
{Wilman} D.~J.,  {Zibetti} S.,   {Budav{\'a}ri} T.,  2010, \mn@doi [\mnras]
  {10.1111/j.1365-2966.2010.16845.x}, \href
  {https://ui.adsabs.harvard.edu/#abs/2010MNRAS.406.1701W} {406, 1701}

\bibitem[\protect\citeauthoryear{{Wisnioski} et~al.,}{{Wisnioski}
  et~al.}{2015}]{wisnioski15}
{Wisnioski} E.,  et~al., 2015, \mn@doi [\apj] {10.1088/0004-637X/799/2/209},
  \href {https://ui.adsabs.harvard.edu/#abs/2015ApJ...799..209W} {799, 209}

\bibitem[\protect\citeauthoryear{{Woo}, {Dekel}, {Faber}  \& {Koo}}{{Woo}
  et~al.}{2015}]{woo15}
{Woo} J.,  {Dekel} A.,  {Faber} S.~M.,   {Koo} D.~C.,  2015, \mn@doi [\mnras]
  {10.1093/mnras/stu2755}, \href
  {https://ui.adsabs.harvard.edu/#abs/2015MNRAS.448..237W} {448, 237}

\bibitem[\protect\citeauthoryear{{Woo}, {Carollo}, {Faber}, {Dekel}  \&
  {Tacchella}}{{Woo} et~al.}{2017}]{woo17}
{Woo} J.,  {Carollo} C.~M.,  {Faber} S.~M.,  {Dekel} A.,   {Tacchella} S.,
  2017, \mn@doi [\mnras] {10.1093/mnras/stw2403}, \href
  {https://ui.adsabs.harvard.edu/#abs/2017MNRAS.464.1077W} {464, 1077}

\bibitem[\protect\citeauthoryear{{Wright} et~al.,}{{Wright}
  et~al.}{2016}]{wright16}
{Wright} A.~H.,  et~al., 2016, \mn@doi [\mnras] {10.1093/mnras/stw832}, \href
  {http://adsabs.harvard.edu/abs/2016MNRAS.460..765W} {460, 765}

\bibitem[\protect\citeauthoryear{{Yang}, {Mo}  \& {van den Bosch}}{{Yang}
  et~al.}{2009}]{yang09}
{Yang} X.,  {Mo} H.~J.,   {van den Bosch} F.~C.,  2009, \mn@doi [\apj]
  {10.1088/0004-637X/695/2/900}, \href
  {http://adsabs.harvard.edu/abs/2009ApJ...695..900Y} {695, 900}

\bibitem[\protect\citeauthoryear{{Zabludoff} \& {Mulchaey}}{{Zabludoff} \&
  {Mulchaey}}{1998}]{zabludoff98}
{Zabludoff} A.~I.,  {Mulchaey} J.~S.,  1998, \mn@doi [\apj] {10.1086/305355},
  \href {https://ui.adsabs.harvard.edu/#abs/1998ApJ...496...39Z} {496, 39}

\bibitem[\protect\citeauthoryear{{Zolotov} et~al.,}{{Zolotov}
  et~al.}{2015}]{zolotov15}
{Zolotov} A.,  et~al., 2015, \mn@doi [\mnras] {10.1093/mnras/stv740}, \href
  {https://ui.adsabs.harvard.edu/#abs/2015MNRAS.450.2327Z} {450, 2327}

\bibitem[\protect\citeauthoryear{{da Cunha}, {Charlot}  \& {Elbaz}}{{da Cunha}
  et~al.}{2008}]{dacunha08}
{da Cunha} E.,  {Charlot} S.,   {Elbaz} D.,  2008, \mn@doi [\mnras]
  {10.1111/j.1365-2966.2008.13535.x}, \href
  {http://adsabs.harvard.edu/abs/2008MNRAS.388.1595D} {388, 1595}

\bibitem[\protect\citeauthoryear{{van Dokkum} \& {Franx}}{{van Dokkum} \&
  {Franx}}{2001}]{vandokkum01}
{van Dokkum} P.~G.,  {Franx} M.,  2001, \mn@doi [\apj] {10.1086/320645}, \href
  {https://ui.adsabs.harvard.edu/#abs/2001ApJ...553...90V} {553, 90}

\bibitem[\protect\citeauthoryear{{van de Sande} et~al.,}{{van de Sande}
  et~al.}{2017a}]{vds17b}
{van de Sande} J.,  et~al., 2017a, \mn@doi [\mnras] {10.1093/mnras/stx1751},
  \href {http://adsabs.harvard.edu/abs/2017MNRAS.472.1272V} {472, 1272}

\bibitem[\protect\citeauthoryear{{van de Sande} et~al.,}{{van de Sande}
  et~al.}{2017b}]{vds17}
{van de Sande} J.,  et~al., 2017b, \mn@doi [\apj]
  {10.3847/1538-4357/835/1/104}, \href
  {http://adsabs.harvard.edu/abs/2017ApJ...835..104V} {835, 104}

\bibitem[\protect\citeauthoryear{{van de Sande} et~al.,}{{van de Sande}
  et~al.}{2018}]{jesse18}
{van de Sande} J.,  et~al., 2018, \mn@doi [Nature Astronomy]
  {10.1038/s41550-018-0436-x}, \href
  {http://adsabs.harvard.edu/abs/2018NatAs...2..483V} {2, 483}

\bibitem[\protect\citeauthoryear{{van den Bergh}}{{van den
  Bergh}}{1976}]{vandberg76}
{van den Bergh} S.,  1976, \apj, \href
  {http://adsabs.harvard.edu/cgi-bin/nph-bib_query?bibcode=1976ApJ...206..883V&db_key=AST}
  {206, 883}

\bibitem[\protect\citeauthoryear{{van den Bosch}, {Aquino}, {Yang}, {Mo},
  {Pasquali}, {McIntosh}, {Weinmann}  \& {Kang}}{{van den Bosch}
  et~al.}{2008}]{vdbosch08}
{van den Bosch} F.~C.,  {Aquino} D.,  {Yang} X.,  {Mo} H.~J.,  {Pasquali} A.,
  {McIntosh} D.~H.,  {Weinmann} S.~M.,   {Kang} X.,  2008, \mn@doi [\mnras]
  {10.1111/j.1365-2966.2008.13230.x}, \href
  {http://adsabs.harvard.edu/abs/2008MNRAS.387...79V} {387, 79}

\bibitem[\protect\citeauthoryear{{van der Wel} et~al.,}{{van der Wel}
  et~al.}{2014}]{vanderwell14}
{van der Wel} A.,  et~al., 2014, \mn@doi [\apj] {10.1088/0004-637X/788/1/28},
  \href {http://adsabs.harvard.edu/abs/2014ApJ...788...28V} {788, 28}

\makeatother
\end{thebibliography}

\newpage

\begin{appendix}
\section{The effect of beam smearing}
\label{appendix}
The typical seeing of SAMI observations is a significant fraction of the effective radius of the targeted galaxies. 
Thus, beam smearing could systematically bias our estimates of the $V/\sigma$ ratio. In this paper, 
we adopted stringent quality cuts ($r_{e}>$2\arcsec and $r_{e}>$2.5$\times$HWHM) to define our 
{\it final sample}, and minimise the effect of beam smearing. However, it is unquestionable that even for 
the {\it final sample}, our estimates of $V/\sigma$ have been systematically lowered by the atmospheric conditions during 
the observations. 

In order to determine if this could affect our main conclusions, here we correct 
the $V/\sigma$ estimates used in this paper for the effect of beam smearing following the empirical recipe recently presented by \cite{graham18}\footnote{See also \cite{harborne19} for an independent test of these corrections.}.
They take advantage of kinematic galaxy models based on the Jeans Anisotropic Modeling method developed by \cite{cappellari08} to derive 
the intrinsic $\lambda_{r}$ parameter ($\lambda_r^{intr}$, \citealp{emsellem07}) of a galaxy from the observed one ($\lambda_{r}^{obs}$). 
This correction is a function of the galaxy's S\'ersic index ($n$), effective radius ($r_{e}$) and the seeing of the observations ($\sigma_{PSF}=FWHM_{PSF}/2.355$):
\begin{equation}
\label{corr1}
\begin{split}
\small
\lambda_{r_{e}}^{obs} = & \lambda_{r_{e}}^{intr} \Big[1+\Big(\frac{\sigma_{PSF}/r_{e}}{0.47}\Big)^{1.76}\Big]^{-0.84} \times \\
& \times\Big[1+(n-2)\Big(0.26\frac{\sigma_{PSF}}{r_e}\Big)\Big]^{-1}
\end{split}
\end{equation} 
Since in this paper we focus on $V/\sigma$, we need to rewrite Eq~\ref{corr1} as a function of $V/\sigma$. 
Following \cite{emsellem07}, we assume 
\begin{equation}
\label{eq2}
\lambda_{r} = \frac{\kappa(V/\sigma)}{\sqrt{1+\kappa^{2}(V/\sigma)^{2}}}
\end{equation} 
For SAMI galaxies, \cite{vds17b} find $\kappa$=0.97 when $V/\sigma$ and $\lambda_{r}$ are measured within one effective 
radius. Thus, we can rewrite Eq.~\ref{eq2} as
\begin{equation}
\label{corr2}
\frac{V}{\sigma} = \frac{\lambda_{r}}{0.97\sqrt{1-(\lambda_{r})^{2}}}
\end{equation} 
If we assume that Eq.~\ref{corr2} is valid for both observed and intrinsic values, 
the effect of beam smearing on $V/\sigma$ is
\begin{equation}
\label{corr3}
\Big(\frac{V}{\sigma}\Big)^{intr}  = \Big(\frac{V}{\sigma}\Big)^{obs}  \frac{\lambda_{r}^{intr} \sqrt{1-(\lambda_{r}^{obs})^{2}}}{\lambda_{r}^{obs} \sqrt{1-(\lambda_{r}^{intr})^{2}}}
\end{equation} 
It is important to note that the last assumption is likely incorrect, as the relation between $\lambda_{r}$ and $V/\sigma$ depends on data quality as well as 
sample selection. In particular, \cite{vds17b} show that $\kappa$ increases slightly with increasing seeing ($\Delta\kappa$=$-$0.02 with a $\Delta FWHM$=0.5-3.0 arcsec seeing) and between different surveys, suggesting that $\kappa$ for the intrinsic value could be higher than for the observed one. Thus, Eq.~\ref{corr3} must be considered as an upper limit to the real effect of beam smearing. This is why in the main paper we prefer to use observed values instead of the corrected ones.

\begin{figure}
\centering
\includegraphics[width=8.5cm]{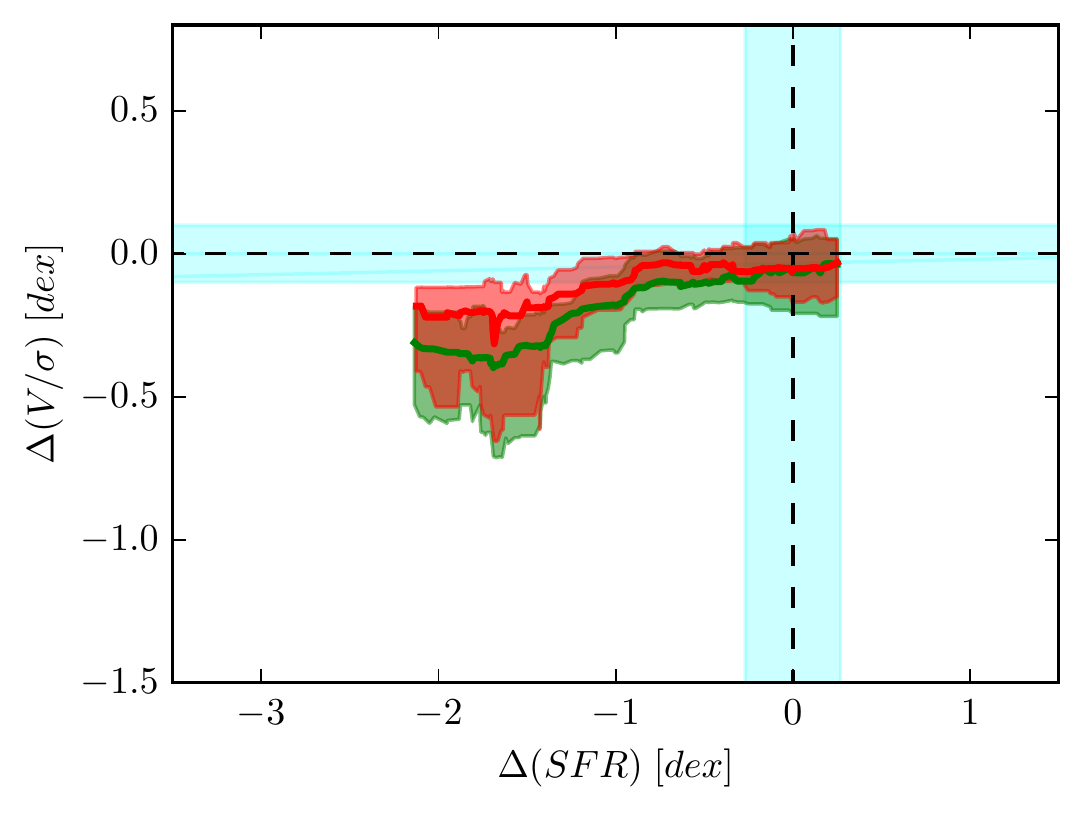}
\caption{Variations in stellar $V/\sigma$ and $SFR$ for satellite galaxies with respect to our control sample of main-sequence, high $V/\sigma$ centrals. 
Dashed lines and cyan bands show the average and standard deviation for the control sample. The green line and shaded region show the running median 
and 20\%-80\% percentile ranges for $\Delta (V/\sigma)$ in bins of $\Delta (SFR)$ for the {\it final sample} used in this paper. The red line and shaded regions 
show how our results would change if we were to apply a beam smearing correction based on the work by Graham et al. (2018).}
\label{beam}
\end{figure}

Fig.~\ref{beam} shows the median and 20\%-80\% percentile ranges of $\Delta (V/\sigma)$ in bins of $\Delta (SFR)$ for the uncorrected (green line; used in the main paper) and corrected (using Eq.~\ref{corr2}; red line) {\it final sample}, respectively. We find that beam smearing has a noticeable effect for galaxies with large negative 
$\Delta (SFR)$, and is almost negligible close to the main sequence. This mainly reflects the difference in apparent size and S\'ersic index between 
passive satellites and star-forming centrals, which translates into a larger correction for passive systems (see Eq.~\ref{corr1}). This shows that, if any, the effect of beam smearing would be to further reduce the change in $\Delta (V/\sigma)$ experienced by satellites during their quenching phase, thus reinforcing the main conclusions of this paper.

\end{appendix}

\end{document}